\documentclass[amsmath,amssymb,11pt,superscriptaddress,reprint, preprintnumbers, notitlepage,aps,twocolumn,nofootinbib]{revtex4-1}
\pdfoutput=1 
\usepackage[utf8]{inputenc}
\usepackage[english]{babel}
\usepackage{amsmath}
\usepackage{graphicx}
\usepackage{dcolumn}
\usepackage{pbox}
\usepackage{amssymb}
\usepackage{epsfig}
\usepackage{slashed}
\usepackage{amssymb}
\usepackage{ mathrsfs }
\usepackage{color}
\usepackage[font=small]{caption}
\usepackage[font=small]{subcaption}
\usepackage{url}
\usepackage{multirow}

\usepackage{lineno}
\usepackage{xcolor,pifont}
\usepackage{comment}

\definecolor{MyLightBlue}{rgb}{0.22,0.51,0.9}
\definecolor{BrickRed}{rgb}{0.8, 0.25, 0.33}
\RequirePackage{hyperref}
\hypersetup{colorlinks, citecolor=blue,linkcolor=red, urlcolor=MyLightBlue}

\makeatletter
\renewcommand\@makecaption[2]{%
  \par
  \vskip\abovecaptionskip
  \begingroup
  
   \small\rmfamily
    \begingroup
     \samepage
     \flushing
     \let\footnote\@footnotemark@gobble
     \@make@capt@title{#1}{#2}\par
    \endgroup
  \endgroup
  \vskip\belowcaptionskip
}
\makeatother

\DeclareUnicodeCharacter{2212}{-}
\begin{document}

\title{\bf Leptoquark-vectorlike quark model for the CDF $m_W$, $(g-2)_\mu$, $R_{K^{(\ast)}}$ anomalies\\ and neutrino mass
}
\author{\bf Talal  Ahmed  Chowdhury}
\email[E-mail: ]{talal@du.ac.bd}
\affiliation{Department of Physics, University of Dhaka, P.O. Box 1000, Dhaka, Bangladesh}
\affiliation{The Abdus Salam International Centre for Theoretical Physics, Strada Costiera 11, I-34014, Trieste, Italy}

\author{\bf Shaikh Saad}
\email[E-mail: ]{shaikh.saad@unibas.ch}
\affiliation{Department of Physics, University of Basel, Klingelbergstrasse\ 82, CH-4056 Basel, Switzerland}

\begin{abstract}
Very recently, a substantial $7\sigma$ deviation  of the $W$-boson mass from the Standard Model (SM) prediction has been reported by the CDF collaboration. Furthermore, the Muon g-2 Experiment recently confirmed the longstanding tension in $(g-2)_\mu$. Besides, the updated result from the LHCb collaboration found evidence
for the breaking of lepton universality in beauty-quark decays, which shows a $3.1\sigma$ discrepancy and is consistent with their previous measurements. Motivated by several of these drawbacks of the SM, in this work, we propose a model consisting of two scalar leptoquarks and a vectorlike quark to simultaneously address the $W$-boson mass shift, the $(g-2)_\mu$, and anomalies in the neutral current transitions of the $B$-meson decays. The proposed model also sheds light on the origin of neutrino mass and can be fully tested at the future colliders.   
\end{abstract}

\maketitle
%%%%%%%%%%%%%%%%%%%%%%%%%%%%%%%%%%%%%%%%%%%%%%%%%%
%%%%%%%%%%%%%%%%%%%%%%%%%%%%%%%%%%%%%%%%%%%%%%%%%%
\section{Introduction}
The $W$-boson mass, $M_W$, is a precisely measured quantity, and even a slight deviation from the predicted value would hint toward physics beyond the Standard Model (SM).  The SM predicts $M_W^{\rm SM} = (80.357 \pm 0.004) \;\textrm{GeV}$, which agrees with the most up-to-date PDG value $M_W^{\rm PDG} = (80.379 \pm 0.012) \;\textrm{GeV}$ at the $2\sigma$ confidence level~\cite{ParticleDataGroup:2020ssz}. Very recently, the CDF collaboration   has reported a new precision measurement of $M_W$ using their
full $8.8$\,fb$^{-1}$ data set that yields~\cite{CDF:2022hxs} 
\begin{align}
M_W^\textrm{CDF-2022}= (80.4335 \pm 0.0094) \;\textrm{GeV},
\end{align}
which deviates from the SM prediction by $7\sigma$, clearly indicating the presence of new physics (NP)~\cite{Fan:2022dck,Zhu:2022tpr,Athron:2022qpo,Du:2022pbp,Yang:2022gvz,deBlas:2022hdk,Tang:2022pxh,Blennow:2022yfm,Zhu:2022scj,Sakurai:2022hwh,Heo:2022dey,Cheung:2022zsb,Lu:2022bgw,Strumia:2022qkt,Fan:2022yly,Cacciapaglia:2022xih,Liu:2022jdq,Lee:2022nqz,Cheng:2022jyi,Song:2022xts,Bagnaschi:2022whn,Paul:2022dds,Bahl:2022xzi,Asadi:2022xiy,DiLuzio:2022xns,Athron:2022isz,Gu:2022htv,Babu:2022pdn,Crivellin:2022fdf,Endo:2022kiw,Han:2022juu,Biekotter:2022abc,Balkin:2022glu,Kawamura:2022uft,Ghoshal:2022vzo,Perez:2022uil,Nagao:2022oin,Kanemura:2022ahw,Heckman:2022the,Ahn:2022xeq,Chowdhury:2022moc,Zeng:2022lkk,Du:2022fqv,Ghorbani:2022vtv,Bhaskar:2022vgk,Baek:2022agi,Cao:2022mif,Borah:2022zim,Batra:2022org,Lee:2022gyf,Cheng:2022aau,Addazi:2022fbj,Heeck:2022fvl,Abouabid:2022lpg,Batra:2022pej,Benbrik:2022dja,Cai:2022cti,Zhou:2022cql,Gupta:2022lrt,Wang:2022dte,Barman:2022qix,Kim:2022hvh,Kim:2022xuo,Dcruz:2022dao,Isaacson:2022rts,Botella:2022rte,He:2021yck,He:2022zjz,Popov:2022ldh}.

The muon's anomalous magnetic moment (AMM) $\Delta a_\mu=(g-2)_\mu/2$, is another quantity that has recently been measured with unprecedented accuracy in the Muon g-2 experiment~\cite{Abi:2021gix}. The result from this experiment is in complete agreement with the previously measured value at  BNL~\cite{Bennett:2006fi}. When these two results are combined, it shows a large  $4.2\sigma$ discrepancy compared to the SM prediction~\cite{Aoyama:2020ynm} (for original works, see Refs.~\cite{Aoyama:2012wk,Aoyama:2019ryr,Czarnecki:2002nt,Gnendiger:2013pva,Davier:2017zfy,Keshavarzi:2018mgv,Colangelo:2018mtw,Hoferichter:2019mqg,Davier:2019can,Keshavarzi:2019abf,Kurz:2014wya,Melnikov:2003xd,Masjuan:2017tvw,Colangelo:2017fiz,Hoferichter:2018kwz,Gerardin:2019vio,Bijnens:2019ghy,Colangelo:2019uex,Blum:2019ugy,Colangelo:2014qya}): 
\begin{align}
&\Delta a_\mu = (2.51\pm 0.59)\times 10^{-9},\label{muongM2}
\end{align}
hinting towards physics beyond the SM (BSM); for a recent review, see e.g.~\cite{Athron:2021iuf}.

In addition, lepton flavor universality (LFU) violating $B$-meson decays have been persistently observed in a series of experiments~\cite{LHCb:2017avl,LHCb:2019hip,Belle:2019oag,BELLE:2019xld,LHCb:2021trn}. The most noteworthy deviation is observed in neutral-current transitions associated with the $R_K-R_{K^*}$ ratios, which are defined as:
\begin{align}
&R_K= \frac{Br\left(B\to K\mu^+\mu^-  \right)}{Br\left(B\to Ke^+e^-  \right)},
\;\;\;
R_{K^{\ast}}= \frac{Br\left(B\to K^{\ast}\mu^+\mu^-  \right)}{Br\left(B\to K^{\ast}e^+e^-  \right)}.
\end{align}

LFU in the SM predicts these ratios to be unity with uncertainties less than $1\%$. However, the most precise measurement by LHCb~\cite{LHCb:2021trn} finds a deficit with a significance of $3.1\sigma$ for $R_K$-ratio. There are several other related observables for which LHCb also found  deficits with respect to
the SM prediction, which are of order $\mathcal{O}(1.5-3.5)\sigma$; for a comprehensive list, see e.g.~\cite{CAMALICH20221}. However, if only the theoretically clean observables: $R_K, R_{K^*}$ ratios and $BR(B_s\to \mu^+\mu^-)$ are taken into account, the data is found to be in $4.2\sigma$  tension with the SM~\cite{Geng:2021nhg}, for a recent review, see~\cite{Angelescu:2021lln}. On the other hand,  when both theoretically clean and dirty observables are considered, global analyses show preferences compared to the SM hypothesis with pulls more than $7\sigma$ (see \cite{Altmannshofer:2021qrr,Geng:2021nhg,Alguero:2021anc,Hurth:2021nsi,Isidori:2021vtc,Kowalska:2019ley,Ciuchini:2021smi,DAmico:2017mtc} for theoretical assumptions and data included in these fits).

On top of these downsides mentioned above, neutrinos remain massless in the SM. On the contrary, several experiments discovered non-zero masses of the neutrinos via observations of neutrino oscillations~\cite{Super-Kamiokande:1998kpq,Super-Kamiokande:2001ljr,SNO:2002tuh,KamLAND:2002uet,KamLAND:2004mhv,K2K:2002icj,MINOS:2006foh}. This work proposes a simultaneous explanation of the $W$-boson mass shift, the tension in the $(g-2)_\mu$, and the anomalies in the neutral current transitions in the $B$-meson decays, as well as neutrino oscillation data. The proposed model employs two scalar leptoquarks (LQs~\cite{Buchmuller:1986zs,Dorsner:2016wpm}): $\widetilde R_2\sim (3,2,1/6)$ and $S_3\sim (\overline 3,3,1/3)$ and a vectorlike quark (VLQ) $\psi\sim (3,2,-5/6)$.  Non-zero mixing between $\widetilde R_2$ and $S_3$ LQs leads to loop corrections to $W$-boson self-energy explaining the CDF anomaly. Utilizing this same mixing, the $(g-2)_\mu$ receives a large NP contribution via the mass flip of the VLQ inside the loop. The $S_3$ LQ, with its interactions with the SM fermions,  addresses the discrepancies in the rare decays of $B$-mesons based on the
neutral current $b\to s\ell\ell$ transitions. Furthermore, non-zero mixing between $\widetilde R_2$ and $S_3$ LQs is also responsible for generating neutrino mass at the one-loop order, and the model put forward in this work can be tested in the ongoing and future experiments.

%%%%%%%%%%%%%%%%%%%%%%%%%%%%%%%%%%%%%%%%%%%%%%%
%%%%%%%%%%%%%%%%%%%%%%%%%%%%%%%%%%%%%%%%%%%%%%%
\section{Proposal}
In this work, we propose a new leptoquark-vectorlike quark model that contains three BSM particles: (i) an iso-doublet  LQ, $\widetilde R_2 (3,2,1/6)$, (ii) an iso-triplet LQ, $S_3 (\overline 3,3,1/3)$, and (iii) an iso-doublet vectorlike quark,  $\psi (3,2,-5/6)$. Here, the quantum numbers are shown under the SM gauge group $SU(3)\times SU(2)\times U(1)$. Furthermore, we assign a baryon number of $1/3$ ($-1/3$) to $\psi, \widetilde R_2$ ($S_3$).
The corresponding component fields of these particles are defined in the following way:
\begin{align}
&\tau .S_3=\begin{pmatrix}
S^{1/3}&\sqrt{2}S^{4/3}\\
\sqrt{2}S^{-2/3}&-S^{1/3}
\end{pmatrix},\\
&\widetilde R_2= \begin{pmatrix}
 R^{2/3}\\ R^{-1/3}
\end{pmatrix},\;\;
\psi= \begin{pmatrix} 
\psi^{-1/3}\\ \psi^{-4/3}
\end{pmatrix}.
\end{align}

%%%%%%%%%%%%%%%%%%%%%%%%%%%%%%%%%%%%%%%%%%%
%%%%%%%%%%%%%%%%%%%%%%%%%%%%%%%%%%%%%%%%%%%
\textbf{Yukawa sector:}--
The relevant part of the Yukawa Lagrangian is given by,
\begin{align}
\mathcal{L_Y}\supset &\; y^S_i\; \overline{Q^c_L}_i \epsilon \left( \tau . S_3\right) L_L +\hat y^\psi_L\; \overline \psi_L \widetilde R_2 \ell_R 
+y^R \overline{d}_R L_L \epsilon \widetilde R_2
\nonumber \\&
+\hat y^\psi_R\; \overline L_L \left( \tau . S_3\right) \psi_R 
+ m_\psi\; \overline \psi_L \psi_R + h.c. \label{yukawa}
\end{align}
One more term is allowed by the gauge symmetries: $y^\prime \overline \psi_L d_R \epsilon H^\ast$. Once the EW symmetry is broken, it generates a mixing between $\psi^{-1/3}$ and $d^{-1/3}$ via $\overline \psi_Ld_R v/\sqrt{2}$, for which we assume the Yukawa coupling to be negligibly small.  Therefore, the mass generation of the SM fermions remain unaltered.   Note that, baryon number assignments as described above forbid two terms: $\mathcal{L_Y}\not\supset  \overline{\psi^c_R} u_{R} \widetilde R_2$ and $\mathcal{L_Y}\not\supset  \overline{Q^c_L} \epsilon \left( \tau_3.S_3 \right) Q_{L}$.

First we focus on a muon-philic scenario, then the Yukawa couplings in Eq.~\eqref{yukawa} take the following form:
\begin{align}
&y^S=\begin{pmatrix}
0&0&0\\0&y^S_{s\mu}&0\\0&y^S_{b\mu}&0
\end{pmatrix}, \;\;\;
\hat y^\psi_L=\begin{pmatrix}
0\\y^\psi_L\\0
\end{pmatrix}, \;\;\;
\hat y^\psi_R=\begin{pmatrix}
0\\y^\psi_R\\0
\end{pmatrix}. \label{yS}
\end{align}
These forms of $\hat y^\psi_{L,R}$ are required to avoid excessive cLFV and provide large NP contribution to $(g-2)_\mu$. The texture of $y^S$ is chosen to explain the anomalies in the neutral current transitions, to be discussed later in the text.  For the simplicity of our work, we take all model parameters to be real.

%%%%%%%%%%%%%%%%%%%%%%%%%%%%%%%%%%%%%%%%%%%
%%%%%%%%%%%%%%%%%%%%%%%%%%%%%%%%%%%%%%%%%%%
\textbf{Scalar sector:}--
The relevant terms in the scalar potential take the following form:
\begin{align}
& V\supset  m^2_R   \widetilde R_2^\dagger \widetilde R_2 +m^2_S S^\dagger_3 S_3
+
\left\{  \mu\; H^\dagger \left( \tau . S_3\right) \widetilde R_2 + h.c. \right\}.  \label{potential}    
\end{align}
The cubic coupling $\mu$ leads to mixing between $\widetilde R_2$ and $S_3$ components, which is crucial in addressing both the $W$-boson mass and  $(g-2)_\mu$ anomalies within the proposed model. Remarkably, the existence of this cubic term also allows neutrinos to have non-zero masses at the one-loop order. In this theory, $\mu$ is one of the most important parameters, and in the limit $\mu\to 0$, one gets $\Delta m_W\to 0$,  $\Delta a_\mu\to 0$, as well as  $m_\nu\to 0$. Note that, to reduce the number of parameters and for the simplicity of our study, scalar quartic couplings are assumed to be somewhat smaller and are not included in Eq.\eqref{potential}. Consequently, all mass splittings are only a function of the trilinear coupling $\mu$.

From the above potential, the mass matrices  in the  $\{S^Q, R^Q\}$ basis are given by,
\begin{align}
M^2_{2/3}=\begin{pmatrix}
m^2_S&\mu v\\
\mu v&m^2_R
\end{pmatrix},\;\;\;
M^2_{1/3}=\begin{pmatrix}
m^2_S&-\mu v/\sqrt{2}\\
-\mu v/\sqrt{2}&m^2_R
\end{pmatrix},
\end{align}
with $v=246$ GeV. 
We denote the weak and mass eigenstates with $X$ and $\hat X$, respectively, which are related by, 
\begin{align}
&S^{\pm Q}=c_x \hat S^{\pm Q} -s_x \hat R^{\pm Q},\\    
&R^{\pm Q}=s_x \hat S^{\pm Q} +c_x \hat R^{\pm Q},
\end{align}
with $x=\theta, \phi$ for $Q=2/3, 1/3$.
Masses and mixing for $\hat X^Q$ states then take the form,
\begin{align}
&m^2_{\hat S_Q, \hat R_Q}=\frac{1}{2}\bigg\{ m^2_S+m^2_R\pm \left[ (m^2_S-m^2_R)^2+ a_Q\mu^2v^2 \right]^{1/2}
\bigg\},
\\&
\sin 2x= \frac{b_Q\mu v}{m^2_{\hat S_Q}-m^2_{\hat R_Q}} \label{theta},
\end{align}
where, $a_Q (b_Q)=4 (2)$ and $2 (-\sqrt{2})$ for $Q=2/3$ and $1/3$, respectively. In the above analysis, we have adopted the convention of $m_S > m_R$.

%%%%%%%%%%%%%%%%%%%%%%%%%%%%%%%%%%%%%%%%%%%%%%%
%%%%%%%%%%%%%%%%%%%%%%%%%%%%%%%%%%%%%%%%%%%%%%%
\textbf{$W$-boson mass shift:}--
The effects of NP phenomena on the electroweak (EW) gauge sector are parameterized in terms of oblique parameters~\cite{Peskin:1990zt,Peskin:1991sw} $S$, $T$, and $U$. Then, the shift in the $W$-boson mass from the NP can be calculated as a function of these oblique parameters~\cite{Grimus:2008nb},
\begin{align}
m_W^2 =  m^2_{W,\rm SM}
\bigg\{
1 + \frac{\alpha_{em}\left(
c_w^2 T -\frac{1}{2} S +
\frac{c_w^2 - s_w^2}{4 s_w^2} U
\right)}{c_w^2 - s_w^2}
\bigg\}.
\label{mW}
\end{align}

When the new CDF data is taken into account in a global electroweak precision fit, the oblique parameters would deviate from their previous (PDG) SM predictions, which several studies have already analyzed~\cite{Lu:2022bgw,Asadi:2022xiy,Bagnaschi:2022whn}, and updated the $2\sigma$ allowed ranges for $S$, $T$, and $U$ parameters in light of the CDF result. By incorporating these new sets of values of oblique parameters in our numerical analysis, we find that for our model with TeV scale LQs, the mass splitting of the mixed LQ states must be of order $\Delta m_{LQ}\sim \mathcal{O}(100)$ GeV to be compatible with the result reported by CDF collaboration.  It is noteworthy to mention that while the CDF II data alone shows a $7\sigma$ deviation; taking the World Average that includes previous measurements (that are compatible with the SM) from the Tevatron and LEP experiments as well as LHC would reduce the tension somewhat (for quantitative analysis, see, for example~\cite{Lu:2022bgw,Asadi:2022xiy,Bagnaschi:2022whn}).

In our model, NP contributions to these parameters originate from the mass splittings among the component fields as a result of mixing between the same charged states from $\widetilde R_2$ and $S_3$. 
We obtain the following one-loop correction to the $T$-parameter for our model,
\begin{align}
\Delta T&=\frac{N_c}{16\pi s^2_W m^2_W} \bigg\{      
\left(s_\phi s_\theta-\sqrt{2} c_\phi c_\theta  \right)^2 \hat F\left[m_{\hat S^{-2/3}},m_{\hat S^{-1/3}}  \right]
\nonumber \\&
+\left(s_\phi c_\theta+\sqrt{2} c_\phi s_\theta  \right)^2 \hat F\left[ m_{\hat R^{-2/3}},m_{\hat S^{-1/3}} \right]
\nonumber \\&
+\left(c_\phi s_\theta+\sqrt{2} s_\phi c_\theta  \right)^2 \hat F\left[m_{\hat S^{-2/3}},m_{\hat R^{-1/3}}  \right]
\nonumber \\&
+\left(c_\phi c_\theta-\sqrt{2} s_\phi s_\theta  \right)^2 \hat F\left[ m_{\hat R^{-2/3}},m_{\hat R^{-1/3}} \right]
\nonumber \\&
+\left(\sqrt{2} c_\phi \right)^2 \hat F\left[m_{\hat S^{1/3}},m_{S^{-4/3}}  \right]
\nonumber \\&
+\left(-\sqrt{2} s_\phi  \right)^2 \hat F\left[m_{\hat R^{1/3}},m_{S^{-4/3}}  \right]\nonumber \\&
-\frac{c_{\phi}^2s_{\phi}^2}{2}\hat F\left[m_{\hat S^{1/3}},m_{\hat R^{-1/3}}\right]-\frac{c_{\theta}^2s_{\theta}^2}{2}\hat F\left[m_{\hat S^{2/3}},m_{\hat R^{-2/3}}\right]
\bigg\},\label{Tpara}
\end{align}
with,
\begin{align}
\hat F(m_1,m_2)= m_1^2+m^2_2-\frac{2m_1^2m^2_2}{m_1^2-m^2_2}\log \left( \frac{m^2_1}{m^2_2} \right).  
\end{align}
In contrast, we find that the NP contribution to $\Delta S$ is
small compared to $\Delta T$, which in our model, takes the
following form:
\begin{align}
\Delta S &= \frac{N_{c}}{\pi m_{Z}^{2}} \bigg\{
-\frac{1}{3}{\cal B}_{22}\left[m_{Z}^2, m^{2}_{S^{4/3}},m^{2}_{S^{4/3}}\right]
\nonumber\\&
+\frac{1}{48}(3+c_{2\theta})(1+3 c_{2\theta}){\cal B}_{22}\left[m_{Z}^2, m^{2}_{\hat{S}^{2/3}},m^{2}_{\hat{S}^{2/3}}\right] \nonumber\\&
+\frac{1}{96}(9-20c_{2\theta}+3c_{4\theta}){\cal B}_{22}\left[m_{Z}^2, m^{2}_{\hat{R}^{2/3}},m^{2}_{\hat{R}^{2/3}}\right]
\nonumber\\&
-\frac{1}{24}(1+3c_{2\phi})s^{2}_{\phi}{\cal B}_{22}\left[m_{Z}^2, m^{2}_{\hat{S}^{1/3}},m^{2}_{\hat{S}^{1/3}}\right]\nonumber\\&
+\frac{1}{24}(-1+3c_{2\phi})c^{2}_{\phi}{\cal B}_{22}\left[m_{Z}^2, m^{2}_{\hat{R}^{1/3}},m^{2}_{\hat{R}^{1/3}}\right]
\nonumber\\&
+\frac{1}{8}s^{2}_{2\theta}{\cal B}_{22}\left[m_{Z}^2, m^{2}_{\hat{S}^{2/3}},m^{2}_{\hat{R}^{2/3}}\right]\nonumber\\&
+\frac{1}{8}s^{2}_{2\phi}{\cal B}_{22}\left[m_{Z}^2, m^{2}_{\hat{S}^{1/3}},m^{2}_{\hat{R}^{1/3}}\right]\bigg\}, \label{Spara}    
\end{align}
here, the expression for the loop function ${\cal B}_{22}(q^2, m_{1}^{2}, m_{2}^{2})$ is defined as~\cite{He:2001tp} (with $x_k=m^2_k/q^2$): 
\begin{align}
&{\cal B}_{22}(q^2, m_{1}^{2}, m_{2}^{2})=  \frac{q^2}{24} \bigg\{
2\ln q^2+\ln(x_1x_2)+
\nonumber\\&
\left[ \left( x_1-x_2 \right)^3-3\left( x_1^2-x_2^2 \right)+ 3\left( x_1-x_2 \right)  \right] \ln \frac{x_1}{x_2} 
\nonumber\\&
-\left[ 2(x_1-x_2)^2 -8(x_1+x_2)+\frac{10}{3} \right]-
\nonumber\\&
\left[ (x_1-x_2)^2-2(x_1+x_2)+1   \right] h(x_1,x_2) -6H(x_1,x_2)
\bigg\},
\end{align}
with
\begin{align}
H(x_1,x_2)=\frac{x_1+x_2}{2}-\frac{x_1x_2}{x_1-x_2}\ln \frac{x_1}{x_2},    
\end{align}
and
\begin{eqnarray}
h(x_1,x_2)&=&\left\{
\begin{array}{ll}
-2\sqrt{\Delta}\left[\tan^{-1}\frac{x_1-x_2+1}{\sqrt{\Delta}}
-\tan^{-1}\frac{x_1-x_2-1}{\sqrt{\Delta}}\right],\\
0,\\
\sqrt{-\Delta}\ln\frac{x_1+x_2-1+\sqrt{-\Delta}}{x_1+x_2-1-\sqrt{-\Delta}},
\end{array}
\right. 
\label{eq:ffun}\\
\Delta&=&2(x_1+x_2)-(x_1-x_2)^2-1  \,,
\end{eqnarray}
for $\Delta>0$, $\Delta=0$, and $\Delta<0$, respectively. For previous works on leptoquarks effects in EW oblique parameters, see, for example, Refs.~\cite{Froggatt:1991qw,Keith:1997fv,Dorsner:2019itg,Dorsner:2020aaz,Crivellin:2020ukd}.

\begin{figure}[th!]
\centering
\includegraphics[width=0.43\textwidth]{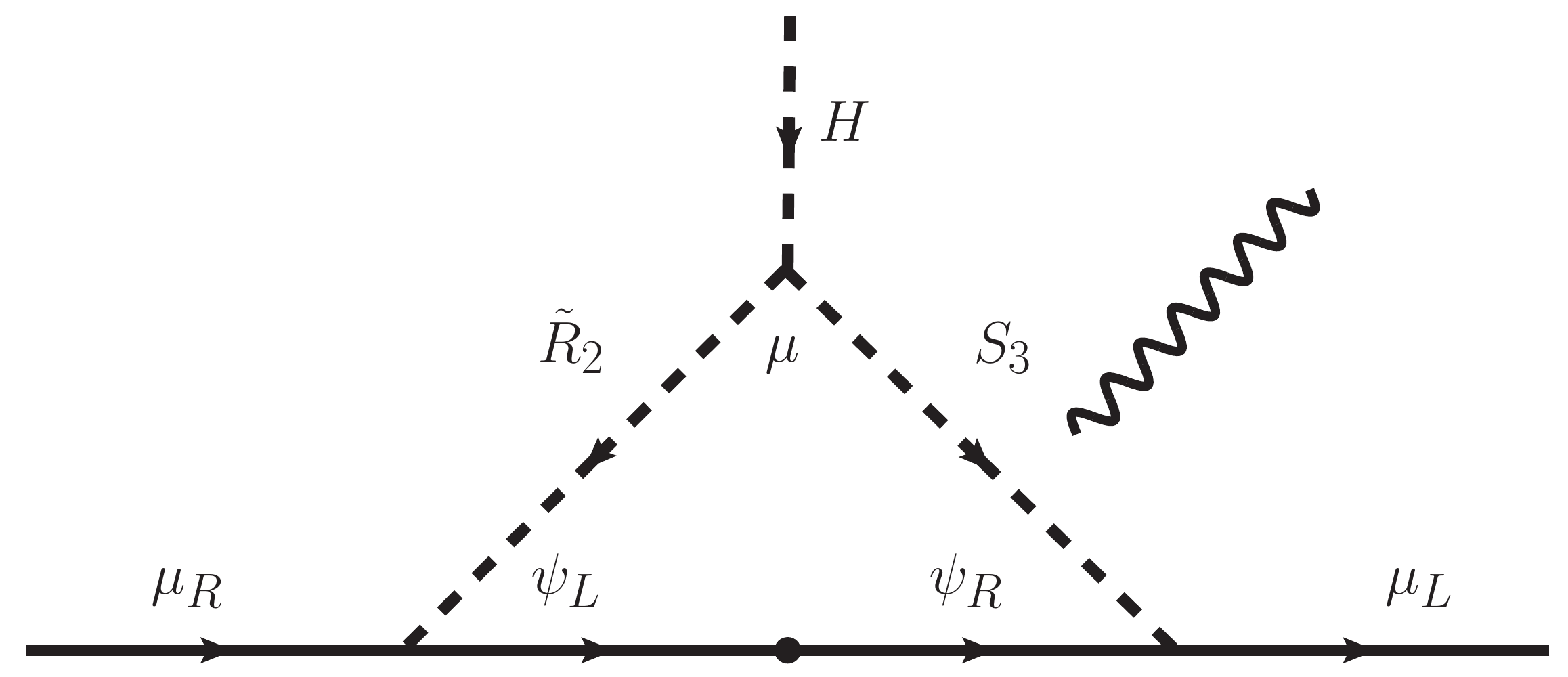}
\caption{Leading order NP contribution to muon AMM (in the weak basis). Photon can be attached to either fermion line or the scalar line.   } \label{fig:AMM}
\end{figure}
%%%%%%%%%%%%%%%%%%%%%%%%%%%%%%%%%%%%%%%%%%%%%%%
%%%%%%%%%%%%%%%%%%%%%%%%%%%%%%%%%%%%%%%%%%%%%%%
\textbf{Muon AMM:}--
In this theory, the AMM of the muon receives NP contributions as shown in Fig.~\ref{fig:AMM}, which can be expressed as~\cite{Leveille:1977rc,Crivellin:2018qmi}, 
\begin{align}
&\Delta a_\mu=- \frac{m_\mu N_c}{4\pi^2} \sum_k \bigg\{ 
\Gamma^{L\ast}_k \Gamma^R_k \frac{m_\psi}{m^2_{\phi_k}} F_k\left( \frac{m^2_\psi}{m^2_{\phi_k}} \right) 
+ 
\nonumber \\&
\left[  |\Gamma^L_k|^2 + |\Gamma^R_k|^2  \right]
\frac{m_\mu}{m^2_{\phi_k}} G_k\left( \frac{m^2_\psi}{m^2_{\phi_k}} \right)
\bigg\},
\end{align}
here sum is taken over $k=\{\hat S^{2/3}, \hat R^{2/3}, \hat S^{1/3}, \hat R^{1/3}\}$ for which we define $\{\Gamma^R_k,\Gamma^L_k\}= \{y^\psi_Ls_\theta,  \sqrt{2}y^\psi_Rc_\theta\}$, $\{y^\psi_Lc_\theta,  -\sqrt{2}y^\psi_Rs_\theta\}$, $\{y^\psi_Ls_\phi,  -y^\psi_Rc_\phi\}$, and $\{y^\psi_Lc_\phi,  y^\psi_Rs_\phi\}$, respectively. While $\hat X^{2/3}$ is propagating in the loop $Q_\psi=-1/3$, and for $\hat X^{-1/3}$ it is   $Q_\psi=-4/3$. And the functions $F(x), G(x)$ are given by,
\begin{align}
&F(x)=f(x)+Q_\psi g(x), \;\;G(x)=\widetilde f(x)+Q_\psi \widetilde g(x),\\
&f(x)=\frac{x^2-1-2 x \log x}{4(x-1)^3},\;\; g(x)=\frac{x-1-\log x}{2(x-1)^2},\\
&\widetilde f(x)=\frac{2x^3+3x^2-6x+1-6x^2\log x}{24(x-1)^4},\;\; \widetilde g(x)=\frac{1}{2}f(x).
\end{align}

By considering only the dominant chirally-enhanced terms, $(g-2)_\mu$ takes the following simpler form:
\begin{align}
\Delta a_\mu&= \frac{3 m_\mu m_\psi y^\psi_Ly^\psi_R}{8 \pi^2} \bigg\{
s_{2\phi} \left[ \frac{F\left( \frac{m^2_\psi}{m^2_{\hat S^{1/3}}} \right)}{m^2_{\hat S^{1/3}}}
-\frac{F\left( \frac{m^2_\psi}{m^2_{\hat R^{1/3}}} \right)}{m^2_{\hat R^{1/3}}}\right]
\nonumber \\&-
\sqrt{2} s_{2\theta} \left[ \frac{F\left( \frac{m^2_\psi}{m^2_{\hat S^{2/3}}} \right)}{m^2_{\hat S^{2/3}}}
-\frac{F\left( \frac{m^2_\psi}{m^2_{\hat R^{2/3}}} \right)}{m^2_{\hat R^{2/3}}}\right] \bigg\}. \label{a_e}
\end{align}

If the external photon leg is removed from the Feynman diagram Fig.~\ref{fig:AMM}, then the corresponding diagram contributes to the muon mass.  Such a chirally-enhanced contribution can in principle generate a large
mass correction $\delta m_\ell$ to the lepton~\cite{Czarnecki:2001pv,Fuyuto:2018scm,Athron:2021iuf}. Theoretically, this can be absorbed by adjusting the pole mass of the lepton, $m^0_\ell$, which, however, introduces large degree of fine-tuning. To avoid fine-tuned solution, in this work, we adopt the criterion $\delta m_\ell \lesssim m^0_\ell$~\cite{Bigaran:2021kmn}.

As aforementioned, for the AMM of the muon, we adopt the theoretical estimate Eq.~\eqref{muongM2} quoted in the 2020 White Paper~\cite{Aoyama:2020ynm}. The SM prediction given in~\cite{Aoyama:2020ynm} is based on the estimate of the leading-order hadronic vacuum
polarization (HVP) contribution ($a^\textrm{hvp}_\mu$), evaluated from a data-driven approach (using dispersion integral involving hadronic cross-section data). This method has an error of $0.6\%$, subject to experimental uncertainties associated with measured cross-section data. On the other hand, lattice QCD calculations for $a^\textrm{hvp}_\mu$, generally face numerous technical challenges. Despite that, a recent lattice computation, namely by the  BMW collaboration~\cite{Borsanyi:2020mff} quotes an error of only  $0.8\%$. If this computation of $a^\textrm{hvp}_\mu$ is considered, then the combined measurement from E989 and E821, when compared to the SM prediction, reduces to $1.5\sigma$ from $4.2\sigma$. Moreover, two more groups,  CLS/Mainz group~\cite{Ce:2022kxy} and ETMC~\cite{Alexandrou:2022amy} just recently released their lattice computations which show consistency with the BMW result. However, the results of~\cite{Ce:2022kxy} is somewhat in tension  with the previous lattice computations for light-quarks by the  RBC/UKQCD collaboration~\cite{RBC:2018dos} and ETMC~\cite{Giusti:2021dvd}. Instead of considering the  HVP from the data-driven method, if the result from ~\cite{Ce:2022kxy} is adopted, the tension between the SM prediction for $(g-2)_\mu$ and experiment would be reduced to $2.9\sigma$. Concerning the other new lattice result of ~\cite{Alexandrou:2022amy}, it agrees with the BMW (CLS/Mainz) group at the level of $1.0\sigma$ ($1.3\sigma$). However, if these new lattice results hold, they point towards a large $\sim 4.2\sigma$ discrepancy with the low-energy $e^+e^-\to$ hadrons cross-section data with respect to Standard Model (SM)
predictions (see also~\cite{Keshavarzi:2020bfy,Crivellin:2020zul,DiLuzio:2021uty,Ce:2022eix}).

\begin{figure}[th!]\centering
\includegraphics[width=0.3\textwidth]{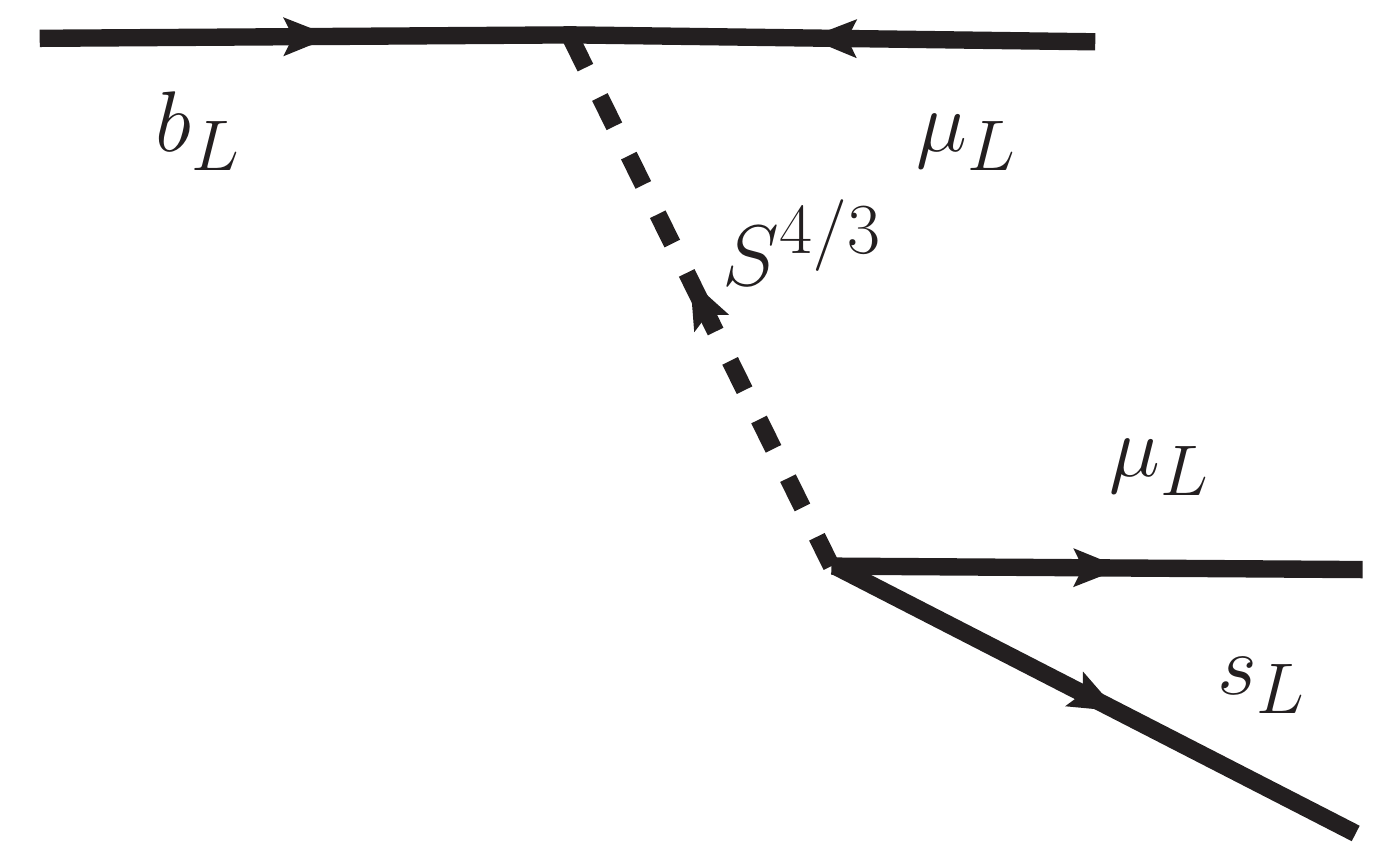}
\caption{Feynman diagram leading to $b\to s\mu^+\mu^-$ transition.} \label{RK}
\end{figure}
%%%%%%%%%%%%%%%%%%%%%%%%%%%%%%%%%%%%%%%%%%%%%%%
%%%%%%%%%%%%%%%%%%%%%%%%%%%%%%%%%%%%%%%%%%%%%%%
\textbf{$R_K-R_{K^\ast}$ anomalies:}--  The effective Hamiltonian responsible for processes of the form $B\to K^{(\ast)} \ell^+ \ell^{\prime -}$ can be described by,
\begin{align}
\mathcal{H}^{dd\ell\ell}_{eff}=-\frac{4 G_F}{\sqrt{2}} V_{tj}V^{\ast}_{ti}\left(\sum_{X=9,10}C_X^{ij,\ell\ell^{\prime}} \mathcal{O}_X^{ij,\ell\ell^{\prime}}\right)  +h.c., \label{Heff} 
\end{align}
where the effective operators are defined as, 
\begin{align}
&\mathcal{O}_9^{ij,\ell\ell^{\prime}}=\frac{\alpha}{4\pi}\left(\overline{d}_i\gamma^{\mu}P_Ld_j\right)\left(\overline{\ell}\gamma_{\mu}\ell^{\prime}\right),\;
\\&
\mathcal{O}_{10}^{ij,\ell\ell^{\prime}}=\frac{\alpha}{4\pi}\left(\overline{d}_i\gamma^{\mu}P_Ld_j\right)\left(\overline{\ell}\gamma_{\mu}\gamma_5\ell^{\prime}\right).
\end{align}

Now, the part of the Lagrangian relevant for $R_{K^{(\ast)}}$-observable is the first term given in Eq~\eqref{yukawa}. We choose to work with the `down-type diagonal' flavor ansatz for which the CKM matrix enters in the interactions associated with the up-type quarks.  Then $S_3$ couplings to SM fermions contain a term of the form:
\begin{align}
\mathcal{L}_{S_3} \supset &
\left( -\sqrt{2}y^S \right)_{ij}   \overline{d^c_L}_i \mu_{Lj} S^{4/3} +\;h.c., \label{S3}
\end{align}
which leads to $b\to s\mu^+\mu^-$ transition as shown in Fig~\ref{RK}. After integrating out the heavy leptoquark and combining the Yukawa part of the Lagrangian associated to $S_3$ as given above, the relevant Wilson coefficients containing in Eq.~\eqref{Heff} at the LQ mass scale generating such neutral current processes take the form, 
\begin{align}
\Delta C_{9}^{\mu\mu}=-\Delta C_{10}^{\mu\mu}=\frac{v^{2}}{V_{t b} V_{t s}^{\ast}}\frac{\pi}{\alpha_{em}} \frac{y^S_{b\mu} \left(y^S_{s \mu} \right)^{*}}{M_{4/3}^{2}}.\label{RK1}
\end{align}
A global fit to the data that includes all $b\to s\mu\mu$ observables,  the $R_{K^{(*)}}$ ratios, and $B_s\to \mu^-\mu^+$ branching ratio prefers  $\Delta C_9^{\mu\mu}=-\Delta C_{10}^{\mu\mu}=-0.39\pm 0.07$ \cite{Altmannshofer:2021qrr} (see also~\cite{Carvunis:2021jga,Geng:2021nhg}).

\begin{figure}[th!]
\centering
\includegraphics[width=0.43\textwidth]{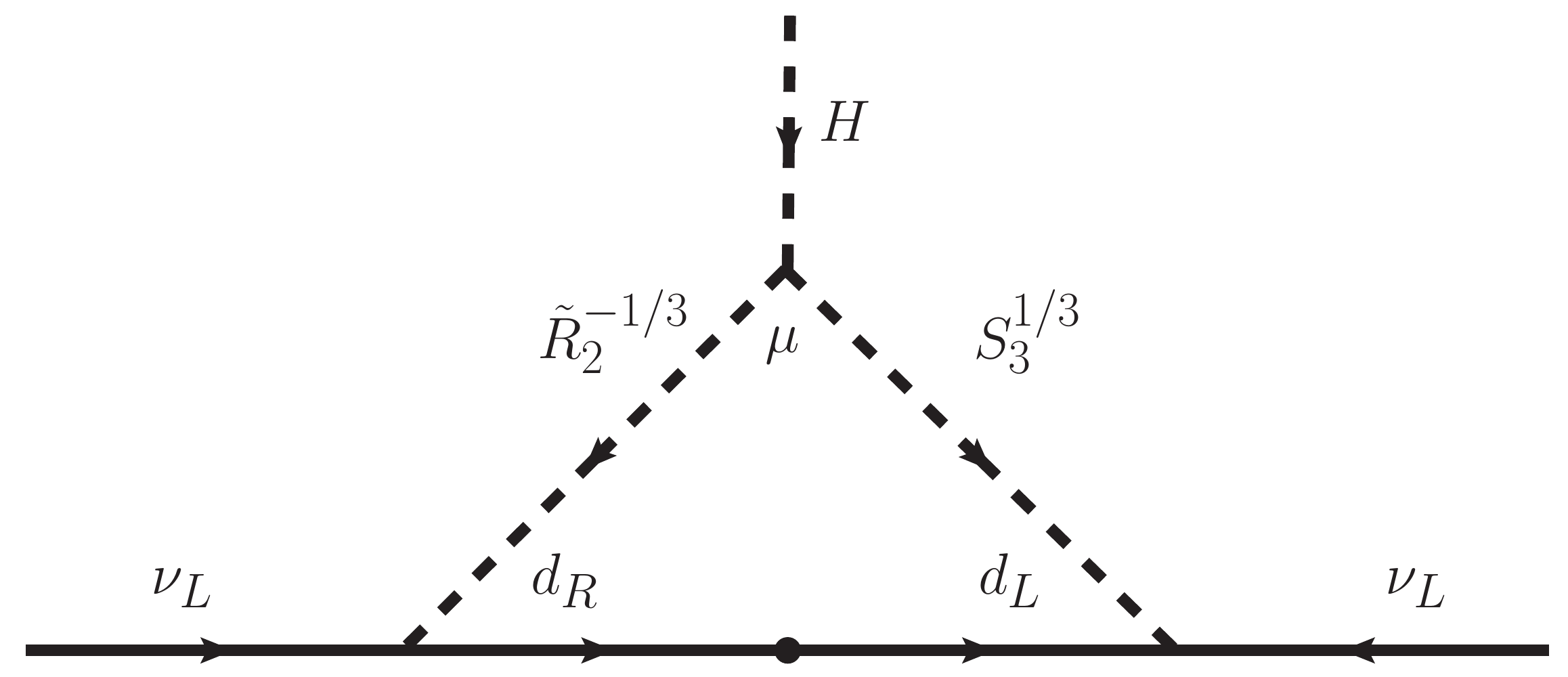}
\caption{Feynman diagram leading to non-zero neutrino masses (in the weak basis).} \label{fig:neutrino}
\end{figure}
%%%%%%%%%%%%%%%%%%%%%%%%%%%%%%%%%%%%%%%%%%%%%%%
%%%%%%%%%%%%%%%%%%%%%%%%%%%%%%%%%%%%%%%%%%%%%%%
\textbf{Neutrino mass:}--
Non-zero mixing between the $\widetilde R_2$ and $S_3$ LQs and BSM Yukawa interactions of the SM fermions with these LQs give rise to neutrino oscillations \cite{Dorsner:2017wwn,Julio:2022ton} in this theory (for LQ effects in leptonic processes, see e.g.~\cite{Crivellin:2020tsz,Crivellin:2020mjs}). Feynman diagram that leads to non-zero neutrino mass is shown in Fig.~\ref{fig:neutrino}, and the neutrino mass formula takes the following form 
\cite{Dorsner:2017wwn}:
\begin{align}
\mathcal{M}^{\nu}_{ij}&=
\frac{\sin 2\phi}{32\pi^2} \sum_{k=d,s,b} m_k
\left[ (y^R)_{ki}(y^S)_{kj} + (y^R)_{kj}(y^S)_{ki} \right]
\nonumber \\& \times 
\left[   
\frac{m^2_{\hat S_{1/3}}\ln{\frac{m^2_k}{m^2_{\hat S_{1/3}}}}}{m^2_{\hat S_{1/3}}-m^2_k}  -
\frac{m^2_{\hat R_{1/3}}\ln{\frac{m^2_k}{m^2_{\hat R_{1/3}}}}}{m^2_{\hat R_{1/3}}-m^2_k}
\right]. \label{nu}
\end{align}
Since down-type quark masses are much smaller than the LQ masses, the above formula can be further simplified, 
\begin{align}
\mathcal{M}^{\nu}\approx \frac{\sin 2\phi}{16\pi^2} \log\left(\frac{m_{\hat R_{1/3}}}{m_{\hat S_{1/3}}}\right) \bigg\{ (y^R)^T m_D y^S + (y^S)^T m_D y^R \bigg\}. \label{numass}
\end{align}

%%%%%%%%%%%%%%%%%%%%%%%%%%%%%%%%%%%%%%%%%%%%%%%
%%%%%%%%%%%%%%%%%%%%%%%%%%%%%%%%%%%%%%%%%%%%%%%
\textbf{Collider constraints:}--
At LHC, $\widetilde R_2$ and $S_3$ LQs can be pair produced~\cite{Diaz:2017lit,Dorsner:2018ynv} via gluon-fusion $pp\to \textrm{LQ}\;\textrm{LQ}^\dagger$. Once produced, each of these LQs would decay to SM fermions. Several searches for LQ pairs have
been made at ATLAS and CMS for different final states with or without neutrinos.  The strongest constraints for our scenario come from decay of these LQs to a third generation quark and a second generation charged lepton, namely $b\mu$ and $t\mu$. For $100\%$ branching ratio to $pp\to b\overline b\mu^+\mu^-$ $(pp\to t\overline t\mu^+\mu^-)$ channel,  LHC provides a lower bound of $m_{\textrm{LQ}}\gtrsim$ 1.7 (1.5) TeV~\cite{ATLAS:2020dsk,ATLAS:2020xov}.  For similar processes with third generation charged lepton, the corresponding bounds are  $m_{\textrm{LQ}}\gtrsim$ 1 TeV $(pp\to b\overline b\tau^+\tau^-)$ and $m_{\textrm{LQ}}\gtrsim$ 1.4 TeV $(pp\to t\overline t\tau^+\tau^-)$, respectively ~\cite{ATLAS:2019qpq,ATLAS:2021oiz}.  

The single production of LQ becomes only relevant for larger Yukawa
couplings to the first and second-generation quarks, which is not the case in our scenario. For a similar reason, non-resonant dipleton searches at the LHC do not provide strong constraints for the parameter space we are interested in.

VLQs can also be pair produced at the LHC through gluon-fusion. If LQs are lighter that VLQs, then each VLQ would mostly decay to a muon and a LQ leading to $pp\to \overline tt\mu^+\mu^-\ell^+\ell^-,\; \overline bb \mu^+\mu^-\ell^+\ell^-$. Processes of these types  have been previously considered  in Ref.~\cite{Dobrescu:2016pda}. We take the LHC bounds on VLQ from~\cite{CMS:2018zkf, ATLAS:2018mpo} that typically correspond to $m_\psi \gtrsim 1.3-1.4$ TeV. 

On the other hand, if LQs are heavier, then pair produced VLQs would still decay to SM fermion final states as mentioned above, via effective 4-fermion operator of the form $y^\psi y^\textrm{LQ}/M^2_\textrm{LQ}\; \left(\overline \psi \ell \right)\left(\ell q \right)$; for details see Ref.~\cite{Dobrescu:2016pda}; where $y^\textrm{LQ}$ and $y^\psi$ represent generic Yukawa couplings of the LQs and VLQ. For TeV scale VLQ with order one couplings,  the decay is prompt even for LQs masses as heavy as 100 TeV. However, if the corresponding LQ couplings with the SM fermions are very small, then the VLQ can become long-lived and form R-hadrons~\cite{Farrar:1978xj}. For such a scenario, by comparing with the limits on production cross-sections for gluinos and squarks, lower bound of $m_\psi \gtrsim 1.5$ TeV  on the VLQ mass is obtained in Ref.~\cite{Criado:2019mvu}.

%%%%%%%%%%%%%%%%%%%%%%%%%%%%%%%%%%%%%%%%%%%%%%%
%%%%%%%%%%%%%%%%%%%%%%%%%%%%%%%%%%%%%%%%%%%%%%%
\textbf{Future collider prospects:}--
Here we point out that   future experiments such as  multi-TeV muon collider (MuC)~\cite{InternationalMuonCollider:2022qki,Jindariani:2022gxj,Aime:2022flm,MuonCollider:2022xlm} and 100 TeV future circular hadron collider (FCC-hh)~\cite{FCC:2018byv,FCC:2018vvp,Bernardi:2022hny} will probe the entire parameter space of the theory relevant for $B$-meson anomaly.  The most efficient way to probe this scenario is via the predominant LQ interactions with the muons since the new physics cannot appear at an arbitrarily high energy scale ($2\to 2$ fermion scattering amplitudes associated to $b\to s\ell\ell$ transitions
saturate the unitarity bound below  80 TeV~\cite{DiLuzio:2017chi}).

From pair-production, FCC-hh will rule out LQ masses almost up to 10 TeV~\cite{Azatov:2022itm}.  On the other hand,  Drell–Yan (DY) $pp\to \mu \overline \mu$ from non-resonant $t$-channel contribution, a large portion of the parameter space in the Yukawa - mass plane will be ruled out~\cite{Azatov:2022itm} leaving part of the parameter space unconstrained (corresponding to $y_{b\mu}\approx y_{s\mu}$ that minimizes the contribution $pp\to \mu \overline \mu$).  

\begin{figure}[b!]
\centering
\includegraphics[width=8.2cm]{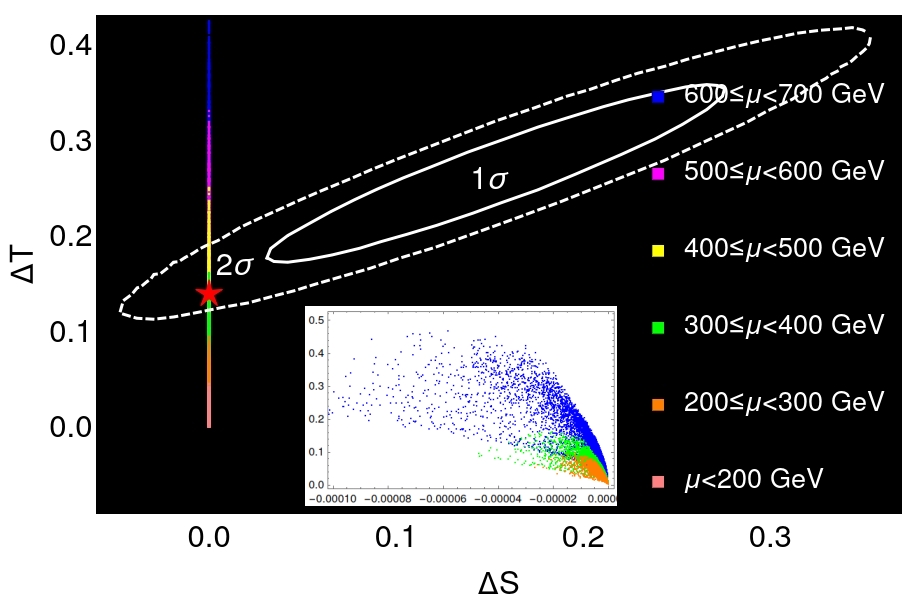}
\caption{Expected values of the $\mu$ parameter to incorporate CDF mass shift of the $W$-boson; see text for details.  The $1\sigma$ and $2\sigma$ regions are obtained from Ref.~\cite{Lu:2022bgw} that performs a global electroweak fit, including the CDF II data. For clarity, part of the parameter space is zoomed in for three different ranges of $\mu$ parameter. The red star represents a particular benchmark point discussed later in the text. } \label{fig:102}
\end{figure}
For muon colliders, the Inverted Drell-Yan (IDY) channel $\mu\overline \mu\to jj$ will constrain the  Yukawa - mass plane, which is similar to FCC-hh scenario (with DY processes). Remarkably, MuC could directly observe an $s$-channel resonance in the   $\mu q\to \mu j$ (due to the quark content inside the muon) for masses up to $\sim s^{1/2}_0$, which would be the most promising on-shell process at muon colliders. When the LQ pair-production, IDY, and $\mu\mu\to\mu j$ processes are combined, MuC10  will probe the entire parameter space~\cite{Azatov:2022itm}.

%%%%%%%%%%%%%%%%%%%%%%%%%%%%%%%%%%%%%%%%%%%%%%%
%%%%%%%%%%%%%%%%%%%%%%%%%%%%%%%%%%%%%%%%%%%%%%%
\textbf{Results and Discussion:}-- 
One of the most important parameters in this model is the scalar cubic coupling $\mu$, which mixes the two LQs. As described above, for $\mu\to 0$, $(g-2)_\mu, \Delta m_W, m_\nu\to 0$.  First, we explicitly demonstrate the required range of $\mu$ to correctly reproduce the electroweak oblique parameters $(\Delta T, \Delta S)$ consistent with the recent CDF II measurement. This is portrayed in Fig.~\ref{fig:102} by randomly varying the LQ mass parameters in the ranges $m_{R}\in (1,5)$ TeV, $m_S-m_R\in (1,500)$ GeV. In making this plot,  $\mu$ is restricted to vary in the range $\mu\lesssim 0.7$ TeV; this upper bound is chosen  here for the sake of clarity and illustration. The $1\sigma$ and $2\sigma$ ranges that favor CDF II data  in the  $(\Delta T, \Delta S)$ plane are taken from Ref.~\cite{Lu:2022bgw} that performed a global fit to the electroweak data. This figure shows that $\mu\gtrsim 350$ GeV is essential to resolve the anomaly.    Note that in our scenario with only the $\mu$ parameter  responsible for splitting the masses of the LQ states, CDF anomaly can only be addressed at the $2\sigma$ confidence limit (C.L.).  It is because the $\Delta S$-parameter given in Eq.~\eqref{Spara} turns out to be always tiny, however, addressing CDF anomaly within  $1\sigma$ C.L. demands $\Delta S\in [0.03, 0.27]$, whereas it is  $\Delta S\in [-0.048, 0.35]$ within $2\sigma$ C.L. In this figure, the red star represents a particular benchmark scenario discussed later in the text.

\begin{figure}[th!]
\centering
\includegraphics[width=8.2cm]{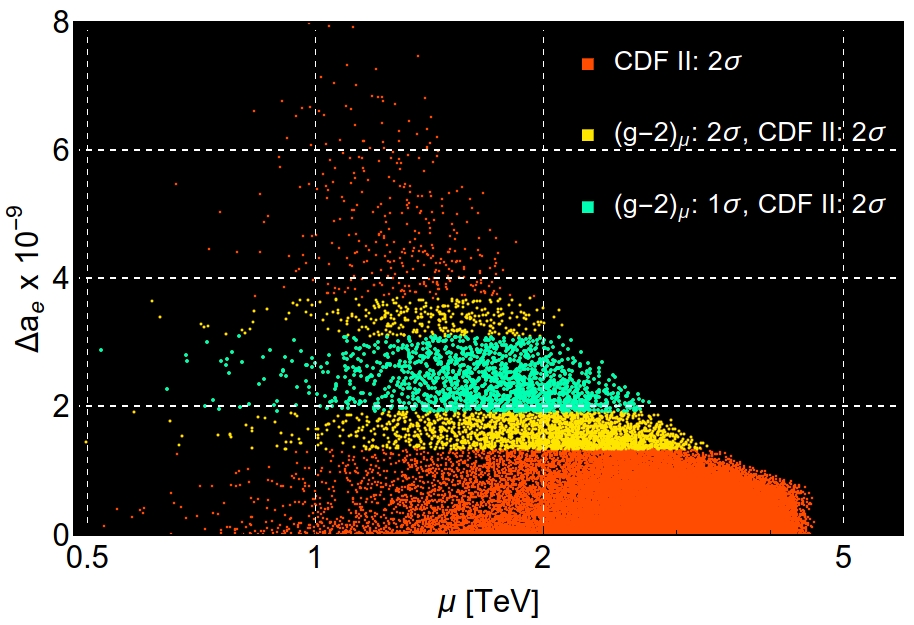}
\caption{Dependence of $(g-2)_\mu$ on the scalar cubic coupling $\mu$. See text for details. } \label{fig:100}
\end{figure}
\begin{figure}[t!]
\centering
\includegraphics[width=8.5cm]{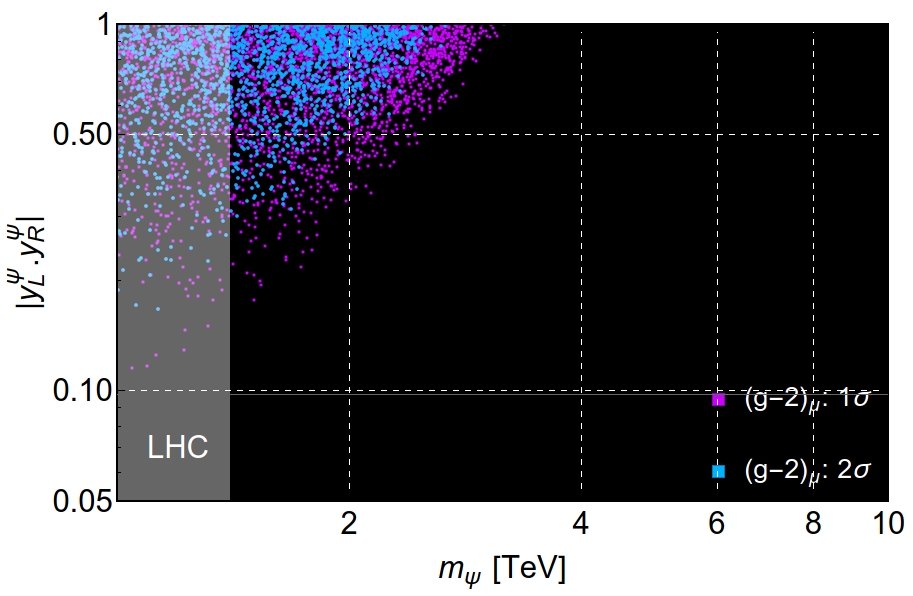}
\caption{Correlations between the product of the Yukawa couplings $|y^\psi_L\cdot y^\psi_R|$  and the mass $m_\psi$ of the VLQ. See text for details. } \label{fig:101}
\end{figure}
\begin{figure}[t!]
\centering
\includegraphics[width=8.5cm]{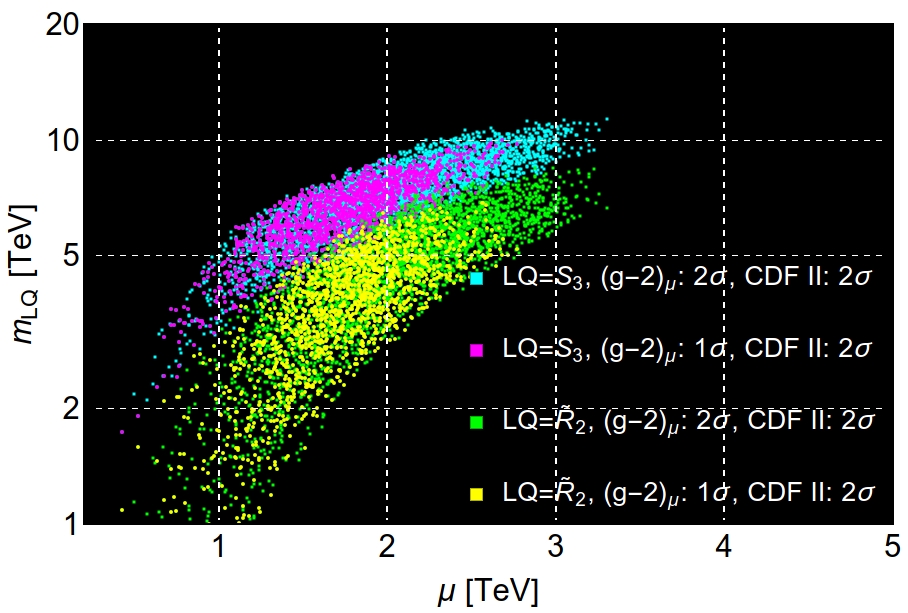}
\caption{Correlations between the $\mu$ parameter  and the masses of the LQs; see text for details. For an example, here we have chosen $Q=2/3$; the other case with $Q=-1/3$ shows indistinguishable behavior. } \label{fig:103}
\end{figure}

The non-trivial functional dependence of $(g-2)_\mu$ on $\mu$ is given in Eq.~\eqref{a_e}, which we graphically  illustrate in Fig.~\ref{fig:100}. Here (for Figs.~\ref{fig:100}, \ref{fig:101}, and \ref{fig:103}), we randomly scan over the relevant parameters in the ranges: $m_{S,R}\in [1,15]$ TeV, $\mu\in [0.1,5]$ TeV,   $m_\psi\in [1,10]$ TeV, and $-y^\psi_L \cdot y^\psi_L\in [10^{-5},1]$. The points in green (yellow) correspond to solutions that simultaneously  satisfy the  $(g-2)_\mu$ anomaly at the $1\sigma$ ($2\sigma$) and the CDF anomaly at the $2\sigma$ C.L. The red points that are consistent with CDF anomaly at the $2\sigma$ C.L, however, fail to reproduce the muon AMM within its $2\sigma$ values.  As can be seen from Fig.~\ref{fig:100}, reproducing correct value of $\Delta a_\mu$ requires $\mu$ in between $\mathcal{O}(0.5)$ to $\mathcal{O}(3)$ TeV. As shown in~\cite{Baker:2021yli}, even though the current LHC measurements~\cite{CMS:2020xwi,ATLAS:2020fzp} of $h\to \mu\mu$ allow a large trilinear coupling, future colliders such as the FCC may be able to measure this coupling and constraint the theory parameter space. 

\begin{figure}[b!]
\centering
\includegraphics[width=8.5cm]{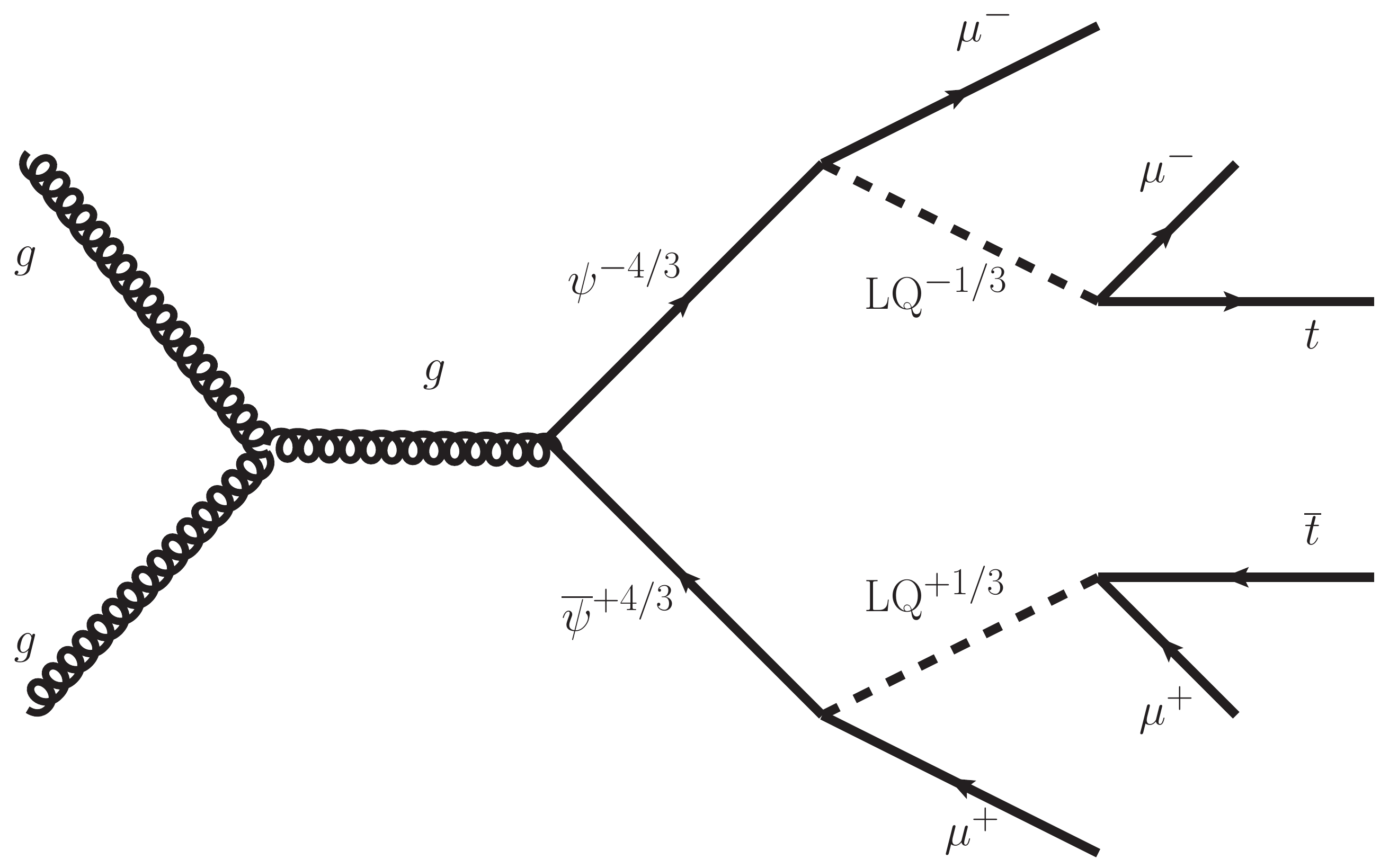}
\caption{Unique signature of the proposed model.} \label{collider}
\end{figure}
Moreover, in Fig.~\ref{fig:101},  we depict the correlations between  the product of the Yukawa couplings $|y^\psi_L\cdot y^\psi_R|$  and  the mass of the VLQ, $m_\psi$. This figure exhibits the required range of the Yukawa couplings as a function of  $m_\psi$. Correctly replicating the measured value of the muon AMM given in Eq.~\eqref{muongM2} calls for a somewhat small mass $m_\psi \lesssim 3$ TeV for the VLQ. Due to these low-mass vectorlike quarks, our model can be distinctively searched for at the LHC. Since the LQs as well as the VLQ predominantly couple to the muons, see Eq.~\eqref{yS}, pair-produced VLQs will lead to a unique signature $pp\to 4\mu+t\overline t$, either via on-shell or off-shell leptoquarks (see discussion above), as shown in Fig.~\ref{collider}. However, a dedicated collider study is beyond the scope of this work.  Branching ratios to other up-type quark final states are expected to be suppressed by CKM elements, and diagrams with neutrinos in the final states are not shown.

Even though both $(g-2)_\mu$ and $W$-boson mass shift heavily depend on the $\mu$-parameter, they are largely insensitive to the scale of the LQ masses. From a naive estimation, $\Delta a_\mu\sim 3m_\mu m_\psi \theta^2_\textrm{LQ} y^2_\textrm{LQ}/(16\pi^2 m^2_\textrm{LQ}) \ln[m^2_\psi/m^2_\textrm{LQ}]$, where a sum over all  LQ contributions must be taken with appropriate signs (the relative signs play important role). Then, for TeV scale VLQ, with order unity Yukawa couplings and maximal mixing, LQ masses as heavy as $m_\textrm{LQ}\sim\mathcal{O}(10)$ TeV still provide correct order $\Delta a_\mu$, as depicted in Fig.~\ref{fig:103}.

Finally, we exemplify how this model can   address the anomaly in the neutral current transitions of the $B$-meson decays as well as incorporate neutrino oscillation data. This, however, requires a careful fit to neutrino observables via minimization of a $\chi^2$-function; a random scan over the parameters is not sufficient.   This is why, we demonstrate the viability of our model with a specific benchmark for which we choose $m_S=2$ TeV, $m_R=1.8$ TeV and mass of the VLQ is  $m_\psi=1.5$ TeV and $\mu=0.62$ TeV. This specific benchmark point along with $y^\psi_Ly^\psi_R=-0.3$ corresponds to $\Delta a_\mu=2.5\times 10^{-9}$ (consistent with Eq.~\eqref{muongM2} at the $1\sigma$ C.L.) and $[\Delta T, \Delta S]=[0.14038,-8\cdot 10^{-6}]$ (this point represents the red star shown in Fig.~\ref{fig:102}).   Note, however, that this fitting procedure is highly non-trivial since the same Yukawa couplings addressing the $R_K-R_{K^\ast}$ anomalies also enter in neutrino observables. In fact, with the texture for $y^S$ with only two non-zero elements as given in Eq.~\eqref{yS}, neutrino masses and mixings utilizing the formula Eq.~\eqref{numass} cannot be accommodated.  To generate viable neutrino masses and mixings, a few more entries must be introduced in $y^S$, which would lead to both charged lepton flavor violation as well as flavor violations in the quark sector~\cite{Saad:2020ucl,Saad:2020ihm,Julio:2022bue} (see also~\cite{Pas:2015hca,Cheung:2016fjo,Deppisch:2016qqd,Cai:2017wry,Guo:2017gxp,Hati:2018fzc, Datta:2019tuj,Popov:2019tyc,Dev:2020qet,Babu:2020hun,Julio:2022ton}).

\begin{figure}[b!]
\centering
\includegraphics[width=8.5cm]{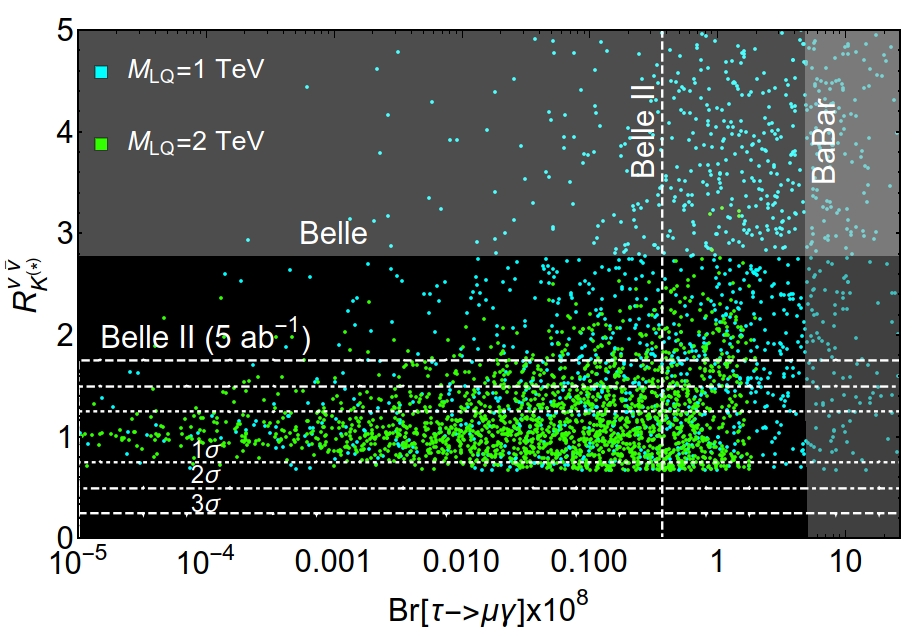}
\caption{The results of random scans showing the correlations between  $BR[\tau\to\mu\gamma]$ and $R^{\nu\overline\nu}_{K^{(*)}}$. Current (future) bound on $BR[\tau\to\mu\gamma]$ from   BaBar Collaboration ~\cite{BaBar:2009hkt} (Super B Factory~\cite{Aushev:2010bq}) is shown by 
shaded light-gray area (vertical dashed white line). For the observable  $R^{\nu\overline\nu}_{K^{(*)}}$, current experimental limit from  Belle collaboration~\cite{Belle:2017oht} is presented by shaded light-gray area.  The regions bounded by dotted, dot-dashed, and dashed white (horizontal) lines depict
the projected reach ($1\pm0.25$) for $R^{\nu\overline\nu}_{K^{(*)}}$ at Belle II \cite{Belle-II:2018jsg} for 5 $ab^{-1}$ of data with $1\sigma$, $2\sigma$, and $3\sigma$ C.L., respectively, assuming the best-fit value is SM-like.    } \label{flavor}
\end{figure}

Our detailed numerical analysis shows that a non-zero $23$-block in $y^S$ is insufficient to satisfy neutrino oscillation data. Therefore, we also introduce a non-zero $31$-entry, which is constrained from $\mu-e$ transition (through CKM rotations).  The most crucial cLFV for the given texture is $\tau\to \mu\gamma$, via which this model may be probed at the upcoming experiment~\cite{Aushev:2010bq}. In addition, various other flavor violating processes are considered in our numerical analysis that includes $Z\to \ell \ell, \ell\to \ell^{'}\ell^{''}\ell^{'''}$, $\tau$ decays into mesons, and several meson decay observables as well as meson-antimeson oscillations; for details, see e.g.~\cite{Saad:2020ucl,Saad:2020ihm}. The observables that provide the most stringent constraints on the model parameters are summarized in Appendix-\ref{A}. As aforementioned, the $23$-block in $y^S$ plays an important role in fitting neutrino observables; some of its entries address the $R_{K^{(*)}}$ anomalies and are required to be sizable. By randomly varying these couplings, interrelations between $BR[\tau\to\mu\gamma]$ and $B\to K^{(*)}\nu\overline \nu$ observables are depicted in Fig.~\ref{flavor} along with their respective experimental bounds.  Clearly,  the prospective measurement of $B\to K^{(*)}\nu\overline\nu$ signal at Belle II experiment shows a promising avenue to test our model.

In addition to the above-mentioned parameters, from a combined fit, we obtain the following Yukawa couplings  addressing $R_K-R_{K^\ast}$ anomalies as well as neutrino oscillation data: 
\begin{align}
y^S&=\left(
\begin{array}{ccc}
 0 & 0 & 0 \\
 0 & 0.01216 & -0.01495 \\
 0.07729 & 0.19480 & 0.07439 \\
\end{array}
\right),
\\
y^R&=10^{-7}    \left(
\begin{array}{ccc}
 1.2583 & 0.29906 & -0.72677 \\
 0.21836 & -0.60107 & -0.03121 \\
 0.12504 & -0.88446 & -1.49160 \\
\end{array}
\right).
\end{align}
This fit corresponds to the following neutrino observables:
\begin{align}
&(m_1,m_2,m_3)=(1.59\times 10^{-2}, 8.60, 50.27)\;\textrm{meV}, 
\\
&(\sin^2\theta_{12},\sin^2\theta_{23},\sin^2\theta_{13})=(0.309, 0.574, 0.02224),
\end{align}
which are in excellent agreement with experimental data and satisfy all flavor constraints.

%%%%%%%%%%%%%%%%%%%%%%%%%%%%%%%%%%%%%%%%%%%%%%%
%%%%%%%%%%%%%%%%%%%%%%%%%%%%%%%%%%%%%%%%%%%%%%%
\textbf{UV completion:}--
Before concluding, we briefly discuss the possible ultraviolet (UV) complete model of the proposed scenario. For demonstration, we choose $SU(5)$ GUT, which is the minimal simple group containing the entire SM gauge group, i.e., $SU(5)\supset SU(3)_c\times SU(2)_L\times U(1)_Y$. As well known, the minimal $SU(5)$ GUT, namely, the Georgi–Glashow model~\cite{Georgi:1974sy} is incompatible with the observed charged fermion masses. To overcome this drawback, Georgi and Jarlskog proposed~\cite{Georgi:1979df} to include a $45_H$. Interestingly, $45_H$ Higgs contains the $S_3$ LQ; $45\supset (3,3,-1/3)$.  On the other hand, $\widetilde R_2$ LQ can be embedded in a new Higgs in the $10_H$ dimensional representation; $10\supset (3,2,1/6)$. Finally, the BSM fermion $\psi$ can emerge from $24_F$ dimensional fermionic representation since $24\supset (3,2,-5/6)+(\overline 3,2,5/6)$.

Within this framework, the Yukawa couplings giving rise to $(g-2)_\mu$ originate from two independent interactions, $10_F 24_F 10^\ast_H$ and $\overline 5_F 24_F 45_H$. Recall that in $SU(5)$ GUT, the SM fermions are embedded in $\overline 5_F+10_F$ representations, and in the Georgi–Glashow model, the scalar $5_H$ contains the SM Higgs. Finally, the cubic coupling of Fig.~\ref{fig:AMM} emanates from the interaction term of the form $10^\ast_H 5_H 45_H$. The Yukawa coupling responsible for addressing the $B$-meson anomaly in the neutral current transition arises from $10_F\overline 5_F 45^\ast_H$ interaction. In addition to the last two interaction terms as aforementioned, neutrino mass generation utilizes another Yukawa coupling with the $\widetilde R_2$ LQ that appears from $\overline 5_F\overline 5_F 10_H$ term in the Lagrangian.

One requires additional fine-tuning on top of the usual doublet-triplet splitting to keep the required multiplets ($S_3, \widetilde R_2, \psi$) light compared to their partners which we assume to reside at the GUT scale. Further tuning of parameters would be required  to keep more states light to achieve successful gauge coupling unification, which we do not attempt to address in this work. For the scalars there are enough number of free parameters in the scalar potential to achieve this, for details, see, for example, Ref.~\cite{Saad:2019vjo}. For the BSM fermion sector, there exist only two free parameters ($\mathcal{L}\supset m_F 24^2_F + \lambda 24_F^2 24_H$) that determine  the mass of all the multiplets. To be precise, $m_{(1,1,0)}= m_F-X$, $m_{(1,3,0)}= m_F-3 X$, $m_{(8,1,0)}= m_F+2 X$, and $m_{(3,2,-5/6)}= m_F-X/2$, where $X=\lambda \;v_\textrm{GUT}/\sqrt{30}$ and   the vev of the Higgs in the adjoint representation is $\langle 24_H\rangle = \textrm{diag}(2,2,2,-3,-3)v_\textrm{GUT}/\sqrt{30}$ that breaks the GUT symmetry to the SM gauge group. It is straightforward to see that in the limit of $X\to 2m_F$ (this leads to an additional fine-tuning on top of the well known doublet-triplet splitting in generic GUTs), only $\psi+c.c.$ remain light, whereas rest of the components of $24_F$ lives close to the GUT scale if the natural choice $m_F\sim v_\textrm{GUT}$ is assumed.

Finally, we comment on proton decay constraints. The $\widetilde R_2$ LQ does not contain di-quark couplings, hence does not lead to proton decay. However, $S_3$ LQ, in general, has both lepton-quark as well as quark-quark couplings leading to proton decay. The latter coupling originates from terms of the form $10_F10_F45_H$; as long as this coupling is forbidden (or highly suppressed) and $S_3$ LQ does not mix with any other fields that contain di-quark couplings,  $S_3$ LQ can be kept light without any conflict with the stringent experimental limits on proton decay. Both of these requirements can be satisfied within this framework, leading to additional fine-tuning. Furthermore, the texture zeros in the Yukawa couplings that we consider may be arranged by imposing flavor symmetries, which, however, is beyond the scope of this work.

%%%%%%%%%%%%%%%%%%%%%%%%%%%%%%%%%%%%%%%%%%%%%%%
%%%%%%%%%%%%%%%%%%%%%%%%%%%%%%%%%%%%%%%%%%%%%%%
\section{Conclusion}
In this work, we proposed a simple new physics scenario to simultaneously address several puzzles that cannot be accounted for by the Standard Model alone. The model is comprised of two scalar leptoquarks and a vectorlike quark. One of the most crucial parameters in this theory is the mixing parameter between the two leptoquarks. This mixing generates neutrino masses via quantum corrections at one loop, provides additional contributions to the $W$-boson mass consistent with recent CDF measurement, and plays a non-trivial role in incorporating the longstanding tension in the muon anomalous magnetic moment. The vectorlike quark, assisted with both the leptoquarks, gives rise to the required sizeable new physics contributions to the $(g-2)_\mu$ via chirally enhanced terms proportional to its mass. Furthermore, the iso-triplet leptoquark is responsible for accounting for the deviations observed persistently in the $R_{K^{(*)}}$ ratios. By performing a numerical analysis, we have illustrated how to consistently resolve all these mysteries mentioned above by keeping flavor violations under control. Moreover, the model is within reach of the current and future upgrades of the LHC and has the potential to be fully probed by the future colliders such as future circular hadron collider and multi-TeV muon collider.

\appendix
%%%%%%%%%%%%%%%%%%%%%%%%%%%%%%%%%%%%%%%%
%%%%%%%%%%%%%%%%%%%%%%%%%%%%%%%%%%%%%%%%
\section{Constraints on model parameters}\label{A}
In this appendix, we provide relevant expressions of all the NP contributions to various flavor violating processes. Since the Yukawa couplings of the $\widetilde R_2$ LQ are small as required for neutrino mass generation, here we focus on the constraints associated with the $S_3$ LQ. Since we have chosen flavor conserving Yukawa couplings of the vectorlike-quark, $\psi$ does not lead to flavor violation.

%%%%%%%%%%%%%%%%%%%%%%%%%%%%%%%%%%%%%%%%
%%%%%%%%%%%%%%%%%%%%%%%%%%%%%%%%%%%%%%%%
\subsection{LFV: $\ell\to \ell^{\prime}\gamma$}

In our model, the dominant LFV process arise from $\ell\to \ell^{\prime}\gamma$. The effective Lagrangian leading to such radiative decays of the charged leptons is given by \cite{Lavoura:2003xp}, 
\begin{align}
\mathcal{L}_{\ell\to \ell^{\prime}\gamma}  =\frac{e}{2} \overline{\ell^{\prime}} i\sigma^{\mu\nu}F_{\mu\nu} \left(  \sigma_L^{\ell\ell^{\prime}}P_L + \sigma_R^{\ell\ell^{\prime}}P_R \right)\ell. 
\end{align}
Then the branching ratios associated to these process are calculated by the following formula \cite{Lavoura:2003xp}:
\begin{align}
Br(\ell\to \ell^{\prime}\gamma) = \frac{\tau_{\ell}\;\alpha\; m^3_{\ell}}{4}  \left( |\sigma^{\ell\ell^{\prime}}_L|^2 + |\sigma^{\ell\ell^{\prime}}_R|^2 \right),
\end{align}
where $\tau_{\ell}$ is the lifetime of the initial state lepton. The   expressions of  $\sigma_{L,R}$  are as follows \cite{Lavoura:2003xp, Dorsner:2016wpm}: 
\begin{align}
&\sigma_{L,S_3}^{if}=\frac{iN_c}{16\pi^2 M^2_3}m_i \left\{ 
(Vy^S)^*_{qf}(Vy^S)_{qi}\frac{-1}{12}+(y^S)^*_{qf}y^S_{qi}\frac{1}{3}
\right\}, 
\\
&\sigma_{R,S_3}^{if}=\frac{iN_c}{16\pi^2 M^2_3}m_f \left\{ 
(Vy^S)^*_{qf}(Vy^S)_{qi}\frac{-1}{12}+(y^S)^*_{qf}y^S_{qi}\frac{1}{3}
\right\}.
\end{align}
Here $V$ is the Cabibbo-Kobayashi-Maskawa (CKM) mixing matrix. The current experimental limits on these processes are~\cite{ TheMEG:2016wtm, Aubert:2009ag}:
\begin{align}
&Br\left(\mu\to e\gamma\right)<4.2\times 10^{-13},\\ 
&Br\left(\tau\to e\gamma\right)<3.3\times 10^{-8}, \\
&Br\left(\tau\to \mu\gamma\right)<4.4\times 10^{-8}.
\end{align}
Among these, $\tau\to\mu\gamma$ provides the most stringent constraint in our scenario, and  the respective future sensitivity is of order $BR[\tau\to\mu\gamma]\sim 10^{-9}$ \cite{Aushev:2010bq}.

%%%%%%%%%%%%%%%%%%%%%%%%%%%%%%%%%%%%%%%%
%%%%%%%%%%%%%%%%%%%%%%%%%%%%%%%%%%%%%%%%
\subsection{Z decays: $Z\to \ell \ell^{\prime}$}
Leptonic decays of the  $Z$-boson receive contributions from the LQs that constraint the Yukawa couplings. These processes are explained with the following effective Lagrangian:
\begin{align}
\delta \mathcal{L}^{Z\to \ell \ell^{\prime}}_{eff}=\frac{g}{\cos \theta_W} \sum_{f,i,j} \overline{f}\gamma^{\mu} \left( g^{ij}_{f_L}P_L + g^{ij}_{f_R}P_R \right) f_j Z_{\mu}.  \end{align}
Here $g$ is the $SU(2)_L$ gauge coupling; $g^{ij}$ are dimensionless couplings  measured with great accuracy  at the LEP   \cite{Tanabashi:2018oca} that provide stringent constraints on the associated Yukawa couplings for a fixed LQ mass. NP contributions to these dimensionless couplings can be expressed as follows \cite{Arnan:2019olv}:
\begin{align}
&Re\left[ \delta g^\ell_{L,R} \right]^{ij} =
\nonumber\\
&  \dfrac{3w^u_{tj}(w^u_{ti})^{\ast}}{16 \pi^2} \Bigg{[} (g_{u_{L,R}}-g_{u_{R,L}})\dfrac{x_t (x_t-1- \log x_t)}{(x_t-1)^2} \Bigg{]}\nonumber \\
&+ \dfrac{x_Z}{16\pi^2} \sum_{q=u,c}w^u_{qj}(w^u_{qi})^{\ast}  \Bigg{[} g_{u_{L,R}}\left( \log x_Z-\frac{1}{6} \right)+\frac{g_{\ell_{L,R}}}{6} \Bigg{]}\nonumber \\
&+\dfrac{x_Z}{16\pi^2} \sum_{q=d,s,b}w^d_{qj}(w^d_{qi})^{\ast}  \Bigg{[} g_{d_{L,R}}\left( \log x_Z-\frac{1}{6} \right)+\frac{g_{\ell_{L,R}}}{6} \Bigg{]}.
\end{align}
For $\delta g_L$ ($\delta g_R$), $w^u_{ij}=-(V^*y^S)_{ij}$, $w^d_{ij}=-\sqrt{2}y^S_{ij}$ ($w^u_{ij}=0$, $w^d_{ij}=0$) for $S_3$ LQ. The LEP collaboration  provides the following limits on these NP contributions~\cite{ALEPH:2005ab}:  
\begin{align}
&Re[\delta g^{ee}_L]\leq 3.0\times 10^{-4},\\ 
&Re[\delta g^{\mu\mu}_L]\leq 1.1\times 10^{-3}, \\
&Re[\delta g^{\tau\tau}_L]\leq 5.8\times 10^{-4}.
\end{align}

%%%%%%%%%%%%%%%%%%%%%%%%%%%%%%%%%%%%%%%%
%%%%%%%%%%%%%%%%%%%%%%%%%%%%%%%%%%%%%%%%
\subsection{$\mu\to e$ conversion} 
The $S_3$ LQ mediates $\mu \to e$ transition in nuclei at the tree-level, and the rate of which can be calculated from the following formula \cite{Kitano:2002mt}:     
\begin{align}
&CR(\mu-e) 
= \frac{\Gamma^{\mu-e} }{\Gamma_\text{capture}(Z)},
\\
&\Gamma^{\mu-e} \,= \,2 \,G_F^2\, 
\left| (2 V^{(p)} + g_{LV}^{(u)} V^{(n)}) g_{LV}^{(u)}\right|^2,
\\
&g_{LV}^{(u)}=\frac{-2v^2}{m_{S_3}^2} \,  
(V^{\ast} y^S)_{u\ell'}\, (V^{\ast}y^S)^*_{u\ell}.
\end{align}
Here, $\Gamma_\text{capture}(Z)$ is the total capture rate for a nucleus with atomic number $Z$, which is $13.07\times 10^6$ $s^{-1}$ for gold, and the corresponding nuclear form factors in units of $m_{\mu}^{5/2}$ are given by   $V^{(p)}=0.0974$,  $V^{(n)}=0.146$ \cite{Kitano:2002mt}. The current  sensitivity implies    \cite{Bertl:2006up}, 
\begin{align}
\text{CR}(\mu-e)<7\times 10^{-13}    
\end{align}
whereas  the future projected sensitivity is expected to make almost four orders
of magnitude improvement over the current limit, i.e.,  $CR(\mu\to e)< 10^{-16}$  \cite{Kurup:2011zza, Cui:2009zz, Chang:2000ac, Adamov:2018vin, Bartoszek:2014mya, Pezzullo:2018fzp, Bonventre:2019grv}.

%%%%%%%%%%%%%%%%%%%%%%%%%%%%%%%%%%%%%%%%
%%%%%%%%%%%%%%%%%%%%%%%%%%%%%%%%%%%%%%%%
\subsection{$B\to K^{(*)}\nu \overline{\nu}$ decays}
Contributions to the left-handed currents in the $b\to s \ell \ell$ process unavoidably imply contributions to $B\to K^{(*)}\nu \overline{\nu}$ decays which are well constrained by experiments. The $S_3$ LQs can induce $B\to K^{(*)}\nu \overline{\nu}$ decay at the tree-level via $d_k\to d_j\nu\overline{\nu}$ processes.  The Wilson coefficient responsible for such decays associated with $b\to s$ transition takes the form,
\begin{align}
C^{fi}_L= \frac{\pi v^2}{2 V_{tb}V^*_{ts}\alpha}  \frac{y^S_{bi}(y^S)^*_{sf}}{M^2_3} .  
\end{align}
Then the branching ratio for $B\to K^{(*)}\nu \overline{\nu}$ can be expressed as~\cite{Buras:2014fpa}:
\begin{align}
R^{\nu \overline{\nu}}_{K^{(*)}}= \frac{1}{3 |C^{\text{SM}}_L|^2} \sum_{i,f=1}^3 \left|  \delta^{fi}C^{\text{SM}}_L +C^{fi}_L \right|^2,   
\end{align}
where  $C^{\text{SM}}_L=-1.47/\sin^2\theta_W$ is the SM contribution. The Belle collaboration  limits these ratios to be  $R^{\nu \overline{\nu}}_K < 3.9$ and $R^{\nu \overline{\nu}}_{K^*} < 2.7$ \cite{Grygier:2017tzo}.

%%%%%%%%%%%%%%%%%%%%%%%%%%%%%%%%%%%%%%%%
%%%%%%%%%%%%%%%%%%%%%%%%%%%%%%%%%%%%%%%%
\subsection{$B^0_s - \overline{B^0_s}$ oscillation} 
$S_3$ contributes to meson-antimeson mixing, and this NP contribution to $B^0_s - \overline{B^0_s}$ mixing can be described by the following effective Lagrangian\cite{Marzocca:2018wcf}:
\begin{align}
\mathcal{L}^{\Delta B=2}_{eff}= -(C^{\text{SM}}_1+C^{NP}_1)  \left( \overline{b}_L \gamma_\mu s_L  \right)^2.   
\end{align}
Here the SM part is $C^{\text{SM}}_1= 2.35/(4\pi^2)\; \left( V_{tb}V^*_{ts} G_F m_W \right)^2$ \cite{Lenz:2010gu} and the NP contribution at the heavy scale ($\Lambda$) is given by \cite{Dorsner:2016wpm,Bobeth:2017ecx,Marzocca:2018wcf,Crivellin:2019dwb},
\begin{align}
C^{NP}_1= \frac{1}{128\pi^2} \frac{5}{M^2_3} \left( \sum_\ell {y^S_{b\ell}}^\ast y^S_{s\ell} \right)^2.
\end{align}
Here we neglect the evolution of $C^{NP}_1$ from high scale to the $m_w$ scale, which is only relevant for precision calculation. Then the mass difference is given by:
\begin{align}
\Delta m_{B_s}^{SM+NP} = \Delta m_{B_s}^{\text{SM}}\left| 1+\frac{C_1^{NP}}{C_1^{\text{SM}}}  \right|,    
\end{align}
where the SM prediction is $\Delta m_{B_s}^{\text{SM}}= (18.3 \pm 2.7)\times 10^{12}s^{-1}$ \cite{Bona:2006sa,Jubb:2016mvq}. This mass difference has been measured in the experiments \cite{Bona:2008jn,Tanabashi:2018oca} with great accuracy, leading to strong constraints on the NP contribution~\cite{UTfit:2007eik}:
\begin{align}
 |C^{NP}_1|\;\; < 2.01\times 10^{-5}\; \textrm{TeV}^{-2}.    
\end{align}

%%%%%%%%%%%%%%%%%%%%%%%%%%%%%%%%%%%%%%%%%%%%%%%
\bibliographystyle{style}
\bibliography{reference}

\providecommand{\href}[2]{#2}\begingroup\raggedright\begin{thebibliography}{100}

\bibitem{ParticleDataGroup:2020ssz}
{\bfseries Particle Data Group} Collaboration, P.~A. Zyla {\em et~al.},
  ``{Review of Particle Physics},''
  \href{http://dx.doi.org/10.1093/ptep/ptaa104}{{\em PTEP} {\bfseries 2020}
  no.~8, (2020) 083C01}.

\bibitem{CDF:2022hxs}
{\bfseries CDF} Collaboration, T.~Aaltonen {\em et~al.}, ``{High-precision
  measurement of the W boson mass with the CDF II detector},''
  \href{http://dx.doi.org/10.1126/science.abk1781}{{\em Science} {\bfseries
  376} no.~6589, (2022) 170--176}.

\bibitem{Fan:2022dck}
Y.-Z. Fan, T.-P. Tang, Y.-L.~S. Tsai, and L.~Wu, ``{Inert Higgs Dark Matter for
  New CDF W-boson Mass and Detection Prospects},''
  \href{http://arxiv.org/abs/2204.03693}{{\ttfamily arXiv:2204.03693
  [hep-ph]}}.

\bibitem{Zhu:2022tpr}
C.-R. Zhu, M.-Y. Cui, Z.-Q. Xia, Z.-H. Yu, X.~Huang, Q.~Yuan, and Y.~Z. Fan,
  ``{GeV antiproton/gamma-ray excesses and the $W$-boson mass anomaly: three
  faces of $\sim 60-70$ GeV dark matter particle?},''
  \href{http://arxiv.org/abs/2204.03767}{{\ttfamily arXiv:2204.03767
  [astro-ph.HE]}}.

\bibitem{Athron:2022qpo}
P.~Athron, A.~Fowlie, C.-T. Lu, L.~Wu, Y.~Wu, and B.~Zhu, ``{The $W$ boson Mass
  and Muon $g-2$: Hadronic Uncertainties or New Physics?},''
  \href{http://arxiv.org/abs/2204.03996}{{\ttfamily arXiv:2204.03996
  [hep-ph]}}.

\bibitem{Du:2022pbp}
X.~K. Du, Z.~Li, F.~Wang, and Y.~K. Zhang, ``{Explaining The Muon $g-2$ Anomaly
  and New CDF II W-Boson Mass in the Framework of (Extra)Ordinary Gauge
  Mediation},'' \href{http://arxiv.org/abs/2204.04286}{{\ttfamily
  arXiv:2204.04286 [hep-ph]}}.

\bibitem{Yang:2022gvz}
J.~M. Yang and Y.~Zhang, ``{Low energy SUSY confronted with new measurements of
  W-boson mass and muon g-2},''
  \href{http://arxiv.org/abs/2204.04202}{{\ttfamily arXiv:2204.04202
  [hep-ph]}}.

\bibitem{deBlas:2022hdk}
J.~de~Blas, M.~Pierini, L.~Reina, and L.~Silvestrini, ``{Impact of the recent
  measurements of the top-quark and W-boson masses on electroweak precision
  fits},'' \href{http://arxiv.org/abs/2204.04204}{{\ttfamily arXiv:2204.04204
  [hep-ph]}}.

\bibitem{Tang:2022pxh}
T.-P. Tang, M.~Abdughani, L.~Feng, Y.-L.~S. Tsai, and Y.-Z. Fan, ``{NMSSM
  neutralino dark matter for $W$-boson mass and muon $g-2$ and the promising
  prospect of direct detection},''
  \href{http://arxiv.org/abs/2204.04356}{{\ttfamily arXiv:2204.04356
  [hep-ph]}}.

\bibitem{Blennow:2022yfm}
M.~Blennow, P.~Coloma, E.~Fern\'andez-Mart\'\i{}nez, and M.~Gonz\'alez-L\'opez,
  ``{Right-handed neutrinos and the CDF II anomaly},''
  \href{http://arxiv.org/abs/2204.04559}{{\ttfamily arXiv:2204.04559
  [hep-ph]}}.

\bibitem{Zhu:2022scj}
B.-Y. Zhu, S.~Li, J.-G. Cheng, R.-L. Li, and Y.-F. Liang, ``{Using gamma-ray
  observation of dwarf spheroidal galaxy to test a dark matter model that can
  interpret the W-boson mass anomaly},''
  \href{http://arxiv.org/abs/2204.04688}{{\ttfamily arXiv:2204.04688
  [astro-ph.HE]}}.

\bibitem{Sakurai:2022hwh}
K.~Sakurai, F.~Takahashi, and W.~Yin, ``{Singlet extensions and W boson mass in
  the light of the CDF II result},''
  \href{http://arxiv.org/abs/2204.04770}{{\ttfamily arXiv:2204.04770
  [hep-ph]}}.

\bibitem{Heo:2022dey}
Y.~Heo, D.-W. Jung, and J.~S. Lee, ``{Impact of the CDF $W$-mass anomaly on two
  Higgs doublet model},'' \href{http://arxiv.org/abs/2204.05728}{{\ttfamily
  arXiv:2204.05728 [hep-ph]}}.

\bibitem{Cheung:2022zsb}
K.~Cheung, W.-Y. Keung, and P.-Y. Tseng, ``{Iso-doublet Vector Leptoquark
  solution to the Muon $g-2$, $R_{K, K^*}$, $R_{D,D^*}$, and $W$-mass
  Anomalies},'' \href{http://arxiv.org/abs/2204.05942}{{\ttfamily
  arXiv:2204.05942 [hep-ph]}}.

\bibitem{Lu:2022bgw}
C.-T. Lu, L.~Wu, Y.~Wu, and B.~Zhu, ``{Electroweak Precision Fit and New
  Physics in light of $W$ Boson Mass},''
  \href{http://arxiv.org/abs/2204.03796}{{\ttfamily arXiv:2204.03796
  [hep-ph]}}.

\bibitem{Strumia:2022qkt}
A.~Strumia, ``{Interpreting electroweak precision data including the $W$-mass
  CDF anomaly},'' \href{http://arxiv.org/abs/2204.04191}{{\ttfamily
  arXiv:2204.04191 [hep-ph]}}.

\bibitem{Fan:2022yly}
J.~Fan, L.~Li, T.~Liu, and K.-F. Lyu, ``{$W$-Boson Mass, Electroweak Precision
  Tests and SMEFT},'' \href{http://arxiv.org/abs/2204.04805}{{\ttfamily
  arXiv:2204.04805 [hep-ph]}}.

\bibitem{Cacciapaglia:2022xih}
G.~Cacciapaglia and F.~Sannino, ``{The W boson mass weighs in on the
  non-standard Higgs},'' \href{http://arxiv.org/abs/2204.04514}{{\ttfamily
  arXiv:2204.04514 [hep-ph]}}.

\bibitem{Liu:2022jdq}
X.~Liu, S.-Y. Guo, B.~Zhu, and Y.~Li, ``{Unifying gravitational waves with $W$
  boson, FIMP dark matter, and Majorana Seesaw mechanism},''
  \href{http://arxiv.org/abs/2204.04834}{{\ttfamily arXiv:2204.04834
  [hep-ph]}}.

\bibitem{Lee:2022nqz}
H.~M. Lee and K.~Yamashita, ``{A Model of Vector-like Leptons for the Muon
  $g-2$ and the $W$ Boson Mass},''
  \href{http://arxiv.org/abs/2204.05024}{{\ttfamily arXiv:2204.05024
  [hep-ph]}}.

\bibitem{Cheng:2022jyi}
Y.~Cheng, X.-G. He, Z.-L. Huang, and M.-W. Li, ``{Type-II Seesaw Triplet Scalar
  and Its VEV Effects on Neutrino Trident Scattering and W mass},''
  \href{http://arxiv.org/abs/2204.05031}{{\ttfamily arXiv:2204.05031
  [hep-ph]}}.

\bibitem{Song:2022xts}
H.~Song, W.~Su, and M.~Zhang, ``{Electroweak Phase Transition in 2HDM under
  Higgs, Z-pole, and W precision measurements},''
  \href{http://arxiv.org/abs/2204.05085}{{\ttfamily arXiv:2204.05085
  [hep-ph]}}.

\bibitem{Bagnaschi:2022whn}
E.~Bagnaschi, J.~Ellis, M.~Madigan, K.~Mimasu, V.~Sanz, and T.~You, ``{SMEFT
  Analysis of $m_{W}$},'' \href{http://arxiv.org/abs/2204.05260}{{\ttfamily
  arXiv:2204.05260 [hep-ph]}}.

\bibitem{Paul:2022dds}
A.~Paul and M.~Valli, ``{Violation of custodial symmetry from W-boson mass
  measurements},'' \href{http://arxiv.org/abs/2204.05267}{{\ttfamily
  arXiv:2204.05267 [hep-ph]}}.

\bibitem{Bahl:2022xzi}
H.~Bahl, J.~Braathen, and G.~Weiglein, ``{New physics effects on the $W$-boson
  mass from a doublet extension of the SM Higgs sector},''
  \href{http://arxiv.org/abs/2204.05269}{{\ttfamily arXiv:2204.05269
  [hep-ph]}}.

\bibitem{Asadi:2022xiy}
P.~Asadi, C.~Cesarotti, K.~Fraser, S.~Homiller, and A.~Parikh, ``{Oblique
  Lessons from the $W$ Mass Measurement at CDF II},''
  \href{http://arxiv.org/abs/2204.05283}{{\ttfamily arXiv:2204.05283
  [hep-ph]}}.

\bibitem{DiLuzio:2022xns}
L.~Di~Luzio, R.~Gr\"ober, and P.~Paradisi, ``{Higgs physics confronts the $M_W$
  anomaly},'' \href{http://arxiv.org/abs/2204.05284}{{\ttfamily
  arXiv:2204.05284 [hep-ph]}}.

\bibitem{Athron:2022isz}
P.~Athron, M.~Bach, D.~H.~J. Jacob, W.~Kotlarski, D.~St\"ockinger, and
  A.~Voigt, ``{Precise calculation of the W boson pole mass beyond the Standard
  Model with FlexibleSUSY},'' \href{http://arxiv.org/abs/2204.05285}{{\ttfamily
  arXiv:2204.05285 [hep-ph]}}.

\bibitem{Gu:2022htv}
J.~Gu, Z.~Liu, T.~Ma, and J.~Shu, ``{Speculations on the W-Mass Measurement at
  CDF},'' \href{http://arxiv.org/abs/2204.05296}{{\ttfamily arXiv:2204.05296
  [hep-ph]}}.

\bibitem{Babu:2022pdn}
K.~S. Babu, S.~Jana, and V.~P. K., ``{Correlating $W$-Boson Mass Shift with
  Muon \textbackslash{}boldmath${g-2}$ in the 2HDM},''
  \href{http://arxiv.org/abs/2204.05303}{{\ttfamily arXiv:2204.05303
  [hep-ph]}}.

\bibitem{Crivellin:2022fdf}
A.~Crivellin, M.~Kirk, T.~Kitahara, and F.~Mescia, ``{Correlating $t\to cZ$ to
  the $W$ Mass and $B$ Physics with Vector-Like Quarks},''
  \href{http://arxiv.org/abs/2204.05962}{{\ttfamily arXiv:2204.05962
  [hep-ph]}}.

\bibitem{Endo:2022kiw}
M.~Endo and S.~Mishima, ``{New physics interpretation of $W$-boson mass
  anomaly},'' \href{http://arxiv.org/abs/2204.05965}{{\ttfamily
  arXiv:2204.05965 [hep-ph]}}.

\bibitem{Han:2022juu}
X.-F. Han, F.~Wang, L.~Wang, J.~M. Yang, and Y.~Zhang, ``{A joint explanation
  of W-mass and muon g-2 in 2HDM},''
  \href{http://arxiv.org/abs/2204.06505}{{\ttfamily arXiv:2204.06505
  [hep-ph]}}.

\bibitem{Biekotter:2022abc}
T.~Biek\"otter, S.~Heinemeyer, and G.~Weiglein, ``{Excesses in the low-mass
  Higgs-boson search and the W-boson mass measurement},''
  \href{http://arxiv.org/abs/2204.05975}{{\ttfamily arXiv:2204.05975
  [hep-ph]}}.

\bibitem{Balkin:2022glu}
R.~Balkin, E.~Madge, T.~Menzo, G.~Perez, Y.~Soreq, and J.~Zupan, ``{On the
  implications of positive W mass shift},''
  \href{http://arxiv.org/abs/2204.05992}{{\ttfamily arXiv:2204.05992
  [hep-ph]}}.

\bibitem{Kawamura:2022uft}
J.~Kawamura, S.~Okawa, and Y.~Omura, ``{$W$ boson mass and muon $g-2$ in a
  lepton portal dark matter model},''
  \href{http://arxiv.org/abs/2204.07022}{{\ttfamily arXiv:2204.07022
  [hep-ph]}}.

\bibitem{Ghoshal:2022vzo}
A.~Ghoshal, N.~Okada, S.~Okada, D.~Raut, Q.~Shafi, and A.~Thapa, ``{Type III
  seesaw with R-parity violation in light of $m_W$ (CDF)},''
  \href{http://arxiv.org/abs/2204.07138}{{\ttfamily arXiv:2204.07138
  [hep-ph]}}.

\bibitem{Perez:2022uil}
P.~F. Perez, H.~H. Patel, and A.~D. Plascencia, ``{On the $W$-mass and New
  Higgs Bosons},'' \href{http://arxiv.org/abs/2204.07144}{{\ttfamily
  arXiv:2204.07144 [hep-ph]}}.

\bibitem{Nagao:2022oin}
K.~I. Nagao, T.~Nomura, and H.~Okada, ``{A model explaining the new CDF II W
  boson mass linking to muon $g-2$ and dark matter},''
  \href{http://arxiv.org/abs/2204.07411}{{\ttfamily arXiv:2204.07411
  [hep-ph]}}.

\bibitem{Kanemura:2022ahw}
S.~Kanemura and K.~Yagyu, ``{Implication of the $W$ boson mass anomaly at CDF
  II in the Higgs triplet model with a mass difference},''
  \href{http://arxiv.org/abs/2204.07511}{{\ttfamily arXiv:2204.07511
  [hep-ph]}}.

\bibitem{Heckman:2022the}
J.~J. Heckman, ``{Extra $W$-Boson Mass from a D3-Brane},''
  \href{http://arxiv.org/abs/2204.05302}{{\ttfamily arXiv:2204.05302
  [hep-ph]}}.

\bibitem{Ahn:2022xeq}
Y.~H. Ahn, S.~K. Kang, and R.~Ramos, ``{Implications of New CDF-II $W$ Boson
  Mass on Two Higgs Doublet Model},''
  \href{http://arxiv.org/abs/2204.06485}{{\ttfamily arXiv:2204.06485
  [hep-ph]}}.

\bibitem{Chowdhury:2022moc}
T.~A. Chowdhury, J.~Heeck, S.~Saad, and A.~Thapa, ``{$W$ boson mass shift and
  muon magnetic moment in the Zee model},''
  \href{http://arxiv.org/abs/2204.08390}{{\ttfamily arXiv:2204.08390
  [hep-ph]}}.

\bibitem{Zeng:2022lkk}
Y.-P. Zeng, C.~Cai, Y.-H. Su, and H.-H. Zhang, ``{Extra boson mix with Z boson
  explaining the mass of W boson},''
  \href{http://arxiv.org/abs/2204.09487}{{\ttfamily arXiv:2204.09487
  [hep-ph]}}.

\bibitem{Du:2022fqv}
M.~Du, Z.~Liu, and P.~Nath, ``{CDF W mass anomaly from a dark sector with a
  Stueckelberg-Higgs portal},''
  \href{http://arxiv.org/abs/2204.09024}{{\ttfamily arXiv:2204.09024
  [hep-ph]}}.

\bibitem{Ghorbani:2022vtv}
K.~Ghorbani and P.~Ghorbani, ``{$W$-Boson Mass Anomaly from Scale Invariant
  2HDM},'' \href{http://arxiv.org/abs/2204.09001}{{\ttfamily arXiv:2204.09001
  [hep-ph]}}.

\bibitem{Bhaskar:2022vgk}
A.~Bhaskar, A.~A. Madathil, T.~Mandal, and S.~Mitra, ``{Combined explanation of
  $W$-mass, muon $g-2$, $R_{K^{(*)}}$ and $R_{D^{(*)}}$ anomalies in a
  singlet-triplet scalar leptoquark model},''
  \href{http://arxiv.org/abs/2204.09031}{{\ttfamily arXiv:2204.09031
  [hep-ph]}}.

\bibitem{Baek:2022agi}
S.~Baek, ``{Implications of CDF $W$-mass and $(g-2)_\mu$ on
  $U(1)_{L_\mu-L_\tau}$ model},''
  \href{http://arxiv.org/abs/2204.09585}{{\ttfamily arXiv:2204.09585
  [hep-ph]}}.

\bibitem{Cao:2022mif}
J.~Cao, L.~Meng, L.~Shang, S.~Wang, and B.~Yang, ``{Interpreting the $W$ mass
  anomaly in the vectorlike quark models},''
  \href{http://arxiv.org/abs/2204.09477}{{\ttfamily arXiv:2204.09477
  [hep-ph]}}.

\bibitem{Borah:2022zim}
D.~Borah, S.~Mahapatra, and N.~Sahu, ``{Singlet-Doublet Fermion Origin of Dark
  Matter, Neutrino Mass and W-Mass Anomaly},''
  \href{http://arxiv.org/abs/2204.09671}{{\ttfamily arXiv:2204.09671
  [hep-ph]}}.

\bibitem{Batra:2022org}
A.~Batra, S.~K.~A., S.~Mandal, and R.~Srivastava, ``{W boson mass in
  Singlet-Triplet Scotogenic dark matter model},''
  \href{http://arxiv.org/abs/2204.09376}{{\ttfamily arXiv:2204.09376
  [hep-ph]}}.

\bibitem{Lee:2022gyf}
S.~Lee, K.~Cheung, J.~Kim, C.-T. Lu, and J.~Song, ``{Status of the
  two-Higgs-doublet model in light of the CDF $m_W$ measurement},''
  \href{http://arxiv.org/abs/2204.10338}{{\ttfamily arXiv:2204.10338
  [hep-ph]}}.

\bibitem{Cheng:2022aau}
Y.~Cheng, X.-G. He, F.~Huang, J.~Sun, and Z.-P. Xing, ``{Dark photon kinetic
  mixing effects for CDF W mass excess},''
  \href{http://arxiv.org/abs/2204.10156}{{\ttfamily arXiv:2204.10156
  [hep-ph]}}.

\bibitem{Addazi:2022fbj}
A.~Addazi, A.~Marciano, A.~P. Morais, R.~Pasechnik, and H.~Yang, ``{CDF II
  $W$-mass anomaly faces first-order electroweak phase transition},''
  \href{http://arxiv.org/abs/2204.10315}{{\ttfamily arXiv:2204.10315
  [hep-ph]}}.

\bibitem{Heeck:2022fvl}
J.~Heeck, ``{W-boson mass in the triplet seesaw model},''
  \href{http://arxiv.org/abs/2204.10274}{{\ttfamily arXiv:2204.10274
  [hep-ph]}}.

\bibitem{Abouabid:2022lpg}
H.~Abouabid, A.~Arhrib, R.~Benbrik, M.~Krab, and M.~Ouchemhou, ``{Is the new
  CDF $M_W$ measurement consistent with the two higgs doublet model?},''
  \href{http://arxiv.org/abs/2204.12018}{{\ttfamily arXiv:2204.12018
  [hep-ph]}}.

\bibitem{Batra:2022pej}
A.~Batra, S.~K. A, S.~Mandal, H.~Prajapati, and R.~Srivastava, ``{CDF-II $W$
  Boson Mass Anomaly in the Canonical Scotogenic Neutrino-Dark Matter Model},''
  \href{http://arxiv.org/abs/2204.11945}{{\ttfamily arXiv:2204.11945
  [hep-ph]}}.

\bibitem{Benbrik:2022dja}
R.~Benbrik, M.~Boukidi, and B.~Manaut, ``{$W$-mass and 96 GeV excess in
  type-III 2HDM},'' \href{http://arxiv.org/abs/2204.11755}{{\ttfamily
  arXiv:2204.11755 [hep-ph]}}.

\bibitem{Cai:2022cti}
C.~Cai, D.~Qiu, Y.-L. Tang, Z.-H. Yu, and H.-H. Zhang, ``{Corrections to
  electroweak precision observables from mixings of an exotic vector boson in
  light of the CDF $W$-mass anomaly},''
  \href{http://arxiv.org/abs/2204.11570}{{\ttfamily arXiv:2204.11570
  [hep-ph]}}.

\bibitem{Zhou:2022cql}
Q.~Zhou and X.-F. Han, ``{The CDF W-mass, muon g-2, and dark matter in a
  $U(1)_{L_\mu-L_\tau}$ model with vector-like leptons},''
  \href{http://arxiv.org/abs/2204.13027}{{\ttfamily arXiv:2204.13027
  [hep-ph]}}.

\bibitem{Gupta:2022lrt}
R.~S. Gupta, ``{Running away from the T-parameter solution to the W mass
  anomaly},'' \href{http://arxiv.org/abs/2204.13690}{{\ttfamily
  arXiv:2204.13690 [hep-ph]}}.

\bibitem{Wang:2022dte}
J.-W. Wang, X.-J. Bi, P.-F. Yin, and Z.-H. Yu, ``{Electroweak dark matter model
  accounting for the CDF $W$-mass anomaly},''
  \href{http://arxiv.org/abs/2205.00783}{{\ttfamily arXiv:2205.00783
  [hep-ph]}}.

\bibitem{Barman:2022qix}
B.~Barman, A.~Das, and S.~Sengupta, ``{New $W$-Boson mass in the light of
  doubly warped braneworld model},''
  \href{http://arxiv.org/abs/2205.01699}{{\ttfamily arXiv:2205.01699
  [hep-ph]}}.

\bibitem{Kim:2022hvh}
J.~Kim, S.~Lee, P.~Sanyal, and J.~Song, ``{CDF $m_W$ and the muon $g-2$ through
  the Higgs-phobic light pseudoscalar in type-X two-Higgs-doublet model},''
  \href{http://arxiv.org/abs/2205.01701}{{\ttfamily arXiv:2205.01701
  [hep-ph]}}.

\bibitem{Kim:2022xuo}
J.~Kim, ``{Compatibility of muon $g-2$, $W$ mass anomaly in type-X 2HDM},''
  \href{http://arxiv.org/abs/2205.01437}{{\ttfamily arXiv:2205.01437
  [hep-ph]}}.

\bibitem{Dcruz:2022dao}
R.~Dcruz and A.~Thapa, ``{$W$ boson mass, dark matter and $(g-2)_\ell$ in
  ScotoZee neutrino mass model},''
  \href{http://arxiv.org/abs/2205.02217}{{\ttfamily arXiv:2205.02217
  [hep-ph]}}.

\bibitem{Isaacson:2022rts}
J.~Isaacson, Y.~Fu, and C.~P. Yuan, ``{ResBos2 and the CDF W Mass
  Measurement},'' \href{http://arxiv.org/abs/2205.02788}{{\ttfamily
  arXiv:2205.02788 [hep-ph]}}.

\bibitem{Botella:2022rte}
F.~J. Botella, F.~Cornet-Gomez, C.~Mir\'o, and M.~Nebot, ``{Muon and electron
  $g-2$ anomalies in a flavor conserving 2HDM with an oblique view on the CDF
  $M_W$ value},'' \href{http://arxiv.org/abs/2205.01115}{{\ttfamily
  arXiv:2205.01115 [hep-ph]}}.

\bibitem{He:2021yck}
S.-P. He, ``{Leptoquark and vectorlike quark extended models as the explanation
  of the muon g-2 anomaly},''
  \href{http://dx.doi.org/10.1103/PhysRevD.105.035017}{{\em Phys. Rev. D}
  {\bfseries 105} no.~3, (2022) 035017},
  \href{http://arxiv.org/abs/2112.13490}{{\ttfamily arXiv:2112.13490
  [hep-ph]}}.

\bibitem{He:2022zjz}
S.-P. He, ``{A leptoquark and vector-like quark extended model for the
  simultaneous explanation of the $W$ boson mass and muon $g-2$ anomalies},''
  \href{http://arxiv.org/abs/2205.02088}{{\ttfamily arXiv:2205.02088
  [hep-ph]}}.

\bibitem{Popov:2022ldh}
O.~Popov and R.~Srivastava, ``{The Triplet Dirac Seesaw in the View of the
  Recent CDF-II W Mass Anomaly},''
  \href{http://arxiv.org/abs/2204.08568}{{\ttfamily arXiv:2204.08568
  [hep-ph]}}.

\bibitem{Abi:2021gix}
{\bfseries Muon g-2} Collaboration, B.~Abi {\em et~al.}, ``{Measurement of the
  Positive Muon Anomalous Magnetic Moment to 0.46~ppm},''
  \href{http://dx.doi.org/10.1103/PhysRevLett.126.141801}{{\em Phys. Rev.
  Lett.} {\bfseries 126} no.~14, (2021) 141801},
  \href{http://arxiv.org/abs/2104.03281}{{\ttfamily arXiv:2104.03281
  [hep-ex]}}.

\bibitem{Bennett:2006fi}
{\bfseries Muon g-2} Collaboration, G.~W. Bennett {\em et~al.}, ``{Final Report
  of the Muon E821 Anomalous Magnetic Moment Measurement at BNL},''
  \href{http://dx.doi.org/10.1103/PhysRevD.73.072003}{{\em Phys. Rev.}
  {\bfseries D73} (2006) 072003},
\href{http://arxiv.org/abs/hep-ex/0602035}{{\ttfamily arXiv:hep-ex/0602035
  [hep-ex]}}.
%%CITATION = HEP-EX/0602035;%%.

\bibitem{Aoyama:2020ynm}
T.~Aoyama {\em et~al.}, ``{The anomalous magnetic moment of the muon in the
  Standard Model},''
  \href{http://dx.doi.org/10.1016/j.physrep.2020.07.006}{{\em Phys. Rept.}
  {\bfseries 887} (2020) 1--166},
  \href{http://arxiv.org/abs/2006.04822}{{\ttfamily arXiv:2006.04822
  [hep-ph]}}.

\bibitem{Aoyama:2012wk}
T.~Aoyama, M.~Hayakawa, T.~Kinoshita, and M.~Nio, ``{Complete Tenth-Order QED
  Contribution to the Muon g-2},''
  \href{http://dx.doi.org/10.1103/PhysRevLett.109.111808}{{\em Phys. Rev.
  Lett.} {\bfseries 109} (2012) 111808},
\href{http://arxiv.org/abs/1205.5370}{{\ttfamily arXiv:1205.5370 [hep-ph]}}.
%%CITATION = ARXIV:1205.5370;%%.

\bibitem{Aoyama:2019ryr}
T.~Aoyama, T.~Kinoshita, and M.~Nio, ``{Theory of the Anomalous Magnetic Moment
  of the Electron},'' \href{http://dx.doi.org/10.3390/atoms7010028}{{\em Atoms}
  {\bfseries 7} no.~1, (2019) 28}.

\bibitem{Czarnecki:2002nt}
A.~Czarnecki, W.~J. Marciano, and A.~Vainshtein, ``{Refinements in electroweak
  contributions to the muon anomalous magnetic moment},''
  \href{http://dx.doi.org/10.1103/PhysRevD.67.073006,
  10.1103/PhysRevD.73.119901}{{\em Phys. Rev.} {\bfseries D67} (2003) 073006},
  \href{http://arxiv.org/abs/hep-ph/0212229}{{\ttfamily arXiv:hep-ph/0212229
  [hep-ph]}}.
[Erratum: Phys. Rev.D73,119901(2006)].
%%CITATION = HEP-PH/0212229;%%.

\bibitem{Gnendiger:2013pva}
C.~Gnendiger, D.~St\"ockinger, and H.~St\"ockinger-Kim, ``{The electroweak
  contributions to $(g-2)_\mu$ after the Higgs boson mass measurement},''
  \href{http://dx.doi.org/10.1103/PhysRevD.88.053005}{{\em Phys. Rev. D}
  {\bfseries 88} (2013) 053005},
  \href{http://arxiv.org/abs/1306.5546}{{\ttfamily arXiv:1306.5546 [hep-ph]}}.

\bibitem{Davier:2017zfy}
M.~Davier, A.~Hoecker, B.~Malaescu, and Z.~Zhang, ``{Reevaluation of the
  hadronic vacuum polarisation contributions to the Standard Model predictions
  of the muon $g-2$ and ${\alpha (m_Z^2)}$ using newest hadronic cross-section
  data},'' \href{http://dx.doi.org/10.1140/epjc/s10052-017-5161-6}{{\em Eur.
  Phys. J. C} {\bfseries 77} no.~12, (2017) 827},
  \href{http://arxiv.org/abs/1706.09436}{{\ttfamily arXiv:1706.09436
  [hep-ph]}}.

\bibitem{Keshavarzi:2018mgv}
A.~Keshavarzi, D.~Nomura, and T.~Teubner, ``{Muon $g-2$ and $\alpha(M_Z^2)$: a
  new data-based analysis},''
  \href{http://dx.doi.org/10.1103/PhysRevD.97.114025}{{\em Phys. Rev. D}
  {\bfseries 97} no.~11, (2018) 114025},
  \href{http://arxiv.org/abs/1802.02995}{{\ttfamily arXiv:1802.02995
  [hep-ph]}}.

\bibitem{Colangelo:2018mtw}
G.~Colangelo, M.~Hoferichter, and P.~Stoffer, ``{Two-pion contribution to
  hadronic vacuum polarization},''
  \href{http://dx.doi.org/10.1007/JHEP02(2019)006}{{\em JHEP} {\bfseries 02}
  (2019) 006}, \href{http://arxiv.org/abs/1810.00007}{{\ttfamily
  arXiv:1810.00007 [hep-ph]}}.

\bibitem{Hoferichter:2019mqg}
M.~Hoferichter, B.-L. Hoid, and B.~Kubis, ``{Three-pion contribution to
  hadronic vacuum polarization},''
  \href{http://dx.doi.org/10.1007/JHEP08(2019)137}{{\em JHEP} {\bfseries 08}
  (2019) 137}, \href{http://arxiv.org/abs/1907.01556}{{\ttfamily
  arXiv:1907.01556 [hep-ph]}}.

\bibitem{Davier:2019can}
M.~Davier, A.~Hoecker, B.~Malaescu, and Z.~Zhang, ``{A new evaluation of the
  hadronic vacuum polarisation contributions to the muon anomalous magnetic
  moment and to $\mathbf{\boldsymbol\alpha(m_Z^2)}$},''
  \href{http://dx.doi.org/10.1140/epjc/s10052-020-7792-2}{{\em Eur. Phys. J. C}
  {\bfseries 80} no.~3, (2020) 241},
  \href{http://arxiv.org/abs/1908.00921}{{\ttfamily arXiv:1908.00921
  [hep-ph]}}. [Erratum: Eur.Phys.J.C 80, 410 (2020)].

\bibitem{Keshavarzi:2019abf}
A.~Keshavarzi, D.~Nomura, and T.~Teubner, ``{$g-2$ of charged leptons, $\alpha
  (M^2_Z)$ , and the hyperfine splitting of muonium},''
  \href{http://dx.doi.org/10.1103/PhysRevD.101.014029}{{\em Phys. Rev. D}
  {\bfseries 101} no.~1, (2020) 014029},
  \href{http://arxiv.org/abs/1911.00367}{{\ttfamily arXiv:1911.00367
  [hep-ph]}}.

\bibitem{Kurz:2014wya}
A.~Kurz, T.~Liu, P.~Marquard, and M.~Steinhauser, ``{Hadronic contribution to
  the muon anomalous magnetic moment to next-to-next-to-leading order},''
  \href{http://dx.doi.org/10.1016/j.physletb.2014.05.043}{{\em Phys. Lett. B}
  {\bfseries 734} (2014) 144--147},
  \href{http://arxiv.org/abs/1403.6400}{{\ttfamily arXiv:1403.6400 [hep-ph]}}.

\bibitem{Melnikov:2003xd}
K.~Melnikov and A.~Vainshtein, ``{Hadronic light-by-light scattering
  contribution to the muon anomalous magnetic moment revisited},''
  \href{http://dx.doi.org/10.1103/PhysRevD.70.113006}{{\em Phys. Rev.}
  {\bfseries D70} (2004) 113006},
\href{http://arxiv.org/abs/hep-ph/0312226}{{\ttfamily arXiv:hep-ph/0312226
  [hep-ph]}}.
%%CITATION = HEP-PH/0312226;%%.

\bibitem{Masjuan:2017tvw}
P.~Masjuan and P.~Sanchez-Puertas, ``{Pseudoscalar-pole contribution to the
  $(g_{\mu}-2)$: a rational approach},''
  \href{http://dx.doi.org/10.1103/PhysRevD.95.054026}{{\em Phys. Rev. D}
  {\bfseries 95} no.~5, (2017) 054026},
  \href{http://arxiv.org/abs/1701.05829}{{\ttfamily arXiv:1701.05829
  [hep-ph]}}.

\bibitem{Colangelo:2017fiz}
G.~Colangelo, M.~Hoferichter, M.~Procura, and P.~Stoffer, ``{Dispersion
  relation for hadronic light-by-light scattering: two-pion contributions},''
  \href{http://dx.doi.org/10.1007/JHEP04(2017)161}{{\em JHEP} {\bfseries 04}
  (2017) 161}, \href{http://arxiv.org/abs/1702.07347}{{\ttfamily
  arXiv:1702.07347 [hep-ph]}}.

\bibitem{Hoferichter:2018kwz}
M.~Hoferichter, B.-L. Hoid, B.~Kubis, S.~Leupold, and S.~P. Schneider,
  ``{Dispersion relation for hadronic light-by-light scattering: pion pole},''
  \href{http://dx.doi.org/10.1007/JHEP10(2018)141}{{\em JHEP} {\bfseries 10}
  (2018) 141},
\href{http://arxiv.org/abs/1808.04823}{{\ttfamily arXiv:1808.04823 [hep-ph]}}.
%%CITATION = ARXIV:1808.04823;%%.

\bibitem{Gerardin:2019vio}
A.~G\'erardin, H.~B. Meyer, and A.~Nyffeler, ``{Lattice calculation of the pion
  transition form factor with $N_f=2+1$ Wilson quarks},''
  \href{http://dx.doi.org/10.1103/PhysRevD.100.034520}{{\em Phys. Rev. D}
  {\bfseries 100} no.~3, (2019) 034520},
  \href{http://arxiv.org/abs/1903.09471}{{\ttfamily arXiv:1903.09471
  [hep-lat]}}.

\bibitem{Bijnens:2019ghy}
J.~Bijnens, N.~Hermansson-Truedsson, and A.~Rodr\'\i{}guez-S\'anchez,
  ``{Short-distance constraints for the HLbL contribution to the muon anomalous
  magnetic moment},''
  \href{http://dx.doi.org/10.1016/j.physletb.2019.134994}{{\em Phys. Lett. B}
  {\bfseries 798} (2019) 134994},
  \href{http://arxiv.org/abs/1908.03331}{{\ttfamily arXiv:1908.03331
  [hep-ph]}}.

\bibitem{Colangelo:2019uex}
G.~Colangelo, F.~Hagelstein, M.~Hoferichter, L.~Laub, and P.~Stoffer,
  ``{Longitudinal short-distance constraints for the hadronic light-by-light
  contribution to $(g-2)_\mu$ with large-$N_c$ Regge models},''
\href{http://arxiv.org/abs/1910.13432}{{\ttfamily arXiv:1910.13432 [hep-ph]}}.
%%CITATION = ARXIV:1910.13432;%%.

\bibitem{Blum:2019ugy}
T.~Blum, N.~Christ, M.~Hayakawa, T.~Izubuchi, L.~Jin, C.~Jung, and C.~Lehner,
  ``{The hadronic light-by-light scattering contribution to the muon anomalous
  magnetic moment from lattice QCD},''
\href{http://arxiv.org/abs/1911.08123}{{\ttfamily arXiv:1911.08123 [hep-lat]}}.
%%CITATION = ARXIV:1911.08123;%%.

\bibitem{Colangelo:2014qya}
G.~Colangelo, M.~Hoferichter, A.~Nyffeler, M.~Passera, and P.~Stoffer,
  ``{Remarks on higher-order hadronic corrections to the muon
  g\ensuremath{-}2},''
  \href{http://dx.doi.org/10.1016/j.physletb.2014.06.012}{{\em Phys. Lett. B}
  {\bfseries 735} (2014) 90--91},
  \href{http://arxiv.org/abs/1403.7512}{{\ttfamily arXiv:1403.7512 [hep-ph]}}.

\bibitem{Athron:2021iuf}
P.~Athron, C.~Bal\'azs, D.~H.~J. Jacob, W.~Kotlarski, D.~St\"ockinger, and
  H.~St\"ockinger-Kim, ``{New physics explanations of $a_\mu$ in light of the
  FNAL muon g-2 measurement},''
  \href{http://dx.doi.org/10.1007/JHEP09(2021)080}{{\em JHEP} {\bfseries 09}
  (2021) 080}, \href{http://arxiv.org/abs/2104.03691}{{\ttfamily
  arXiv:2104.03691 [hep-ph]}}.

\bibitem{LHCb:2017avl}
{\bfseries LHCb} Collaboration, R.~Aaij {\em et~al.}, ``{Test of lepton
  universality with $B^{0} \rightarrow K^{*0}\ell^{+}\ell^{-}$ decays},''
  \href{http://dx.doi.org/10.1007/JHEP08(2017)055}{{\em JHEP} {\bfseries 08}
  (2017) 055}, \href{http://arxiv.org/abs/1705.05802}{{\ttfamily
  arXiv:1705.05802 [hep-ex]}}.

\bibitem{LHCb:2019hip}
{\bfseries LHCb} Collaboration, R.~Aaij {\em et~al.}, ``{Search for
  lepton-universality violation in $B^+\to K^+\ell^+\ell^-$ decays},''
  \href{http://dx.doi.org/10.1103/PhysRevLett.122.191801}{{\em Phys. Rev.
  Lett.} {\bfseries 122} no.~19, (2019) 191801},
  \href{http://arxiv.org/abs/1903.09252}{{\ttfamily arXiv:1903.09252
  [hep-ex]}}.

\bibitem{Belle:2019oag}
{\bfseries Belle} Collaboration, A.~Abdesselam {\em et~al.}, ``{Test of
  Lepton-Flavor Universality in ${B\to K^\ast\ell^+\ell^-}$ Decays at Belle},''
  \href{http://dx.doi.org/10.1103/PhysRevLett.126.161801}{{\em Phys. Rev.
  Lett.} {\bfseries 126} no.~16, (2021) 161801},
  \href{http://arxiv.org/abs/1904.02440}{{\ttfamily arXiv:1904.02440
  [hep-ex]}}.

\bibitem{BELLE:2019xld}
{\bfseries BELLE} Collaboration, S.~Choudhury {\em et~al.}, ``{Test of lepton
  flavor universality and search for lepton flavor violation in $B \rightarrow
  K\ell \ell$ decays},'' \href{http://dx.doi.org/10.1007/JHEP03(2021)105}{{\em
  JHEP} {\bfseries 03} (2021) 105},
  \href{http://arxiv.org/abs/1908.01848}{{\ttfamily arXiv:1908.01848
  [hep-ex]}}.

\bibitem{LHCb:2021trn}
{\bfseries LHCb} Collaboration, R.~Aaij {\em et~al.}, ``{Test of lepton
  universality in beauty-quark decays},''
  \href{http://dx.doi.org/10.1038/s41567-021-01478-8}{{\em Nature Phys.}
  {\bfseries 18} no.~3, (2022) 277--282},
  \href{http://arxiv.org/abs/2103.11769}{{\ttfamily arXiv:2103.11769
  [hep-ex]}}.

\bibitem{CAMALICH20221}
J.~M. Camalich and M.~Patel, ``New lepton non-universal forces in flavor
  physics?,''
  \href{http://dx.doi.org/https://doi.org/10.1016/j.scib.2021.09.012}{{\em
  Science Bulletin} {\bfseries 67} no.~1, (2022) 1--4}.
  \url{https://www.sciencedirect.com/science/article/pii/S2095927321006307}.

\bibitem{Geng:2021nhg}
L.-S. Geng, B.~Grinstein, S.~J\"ager, S.-Y. Li, J.~Martin~Camalich, and R.-X.
  Shi, ``{Implications of new evidence for lepton-universality violation in
  $b\to s\ell^-\ell^+$- decays},''
  \href{http://dx.doi.org/10.1103/PhysRevD.104.035029}{{\em Phys. Rev. D}
  {\bfseries 104} no.~3, (2021) 035029},
  \href{http://arxiv.org/abs/2103.12738}{{\ttfamily arXiv:2103.12738
  [hep-ph]}}.

\bibitem{Angelescu:2021lln}
A.~Angelescu, D.~Be\v{c}irevi\'c, D.~A. Faroughy, F.~Jaffredo, and
  O.~Sumensari, ``{Single leptoquark solutions to the B-physics anomalies},''
  \href{http://dx.doi.org/10.1103/PhysRevD.104.055017}{{\em Phys. Rev. D}
  {\bfseries 104} no.~5, (2021) 055017},
  \href{http://arxiv.org/abs/2103.12504}{{\ttfamily arXiv:2103.12504
  [hep-ph]}}.

\bibitem{Altmannshofer:2021qrr}
W.~Altmannshofer and P.~Stangl, ``{New Physics in Rare B Decays after Moriond
  2021},'' \href{http://arxiv.org/abs/2103.13370}{{\ttfamily arXiv:2103.13370
  [hep-ph]}}.

\bibitem{Alguero:2021anc}
M.~Alguer\'o, B.~Capdevila, S.~Descotes-Genon, J.~Matias, and M.~Novoa-Brunet,
  ``{$b\rightarrow s\ell ^+\ell ^-$ global fits after $R_{K_S}$ and
  $R_{K^{*+}}$},''
  \href{http://dx.doi.org/10.1140/epjc/s10052-022-10231-1}{{\em Eur. Phys. J.
  C} {\bfseries 82} no.~4, (2022) 326},
  \href{http://arxiv.org/abs/2104.08921}{{\ttfamily arXiv:2104.08921
  [hep-ph]}}.

\bibitem{Hurth:2021nsi}
T.~Hurth, F.~Mahmoudi, D.~M. Santos, and S.~Neshatpour, ``{More indications for
  lepton nonuniversality in $b\to s \ell^-\ell^+$},''
  \href{http://dx.doi.org/10.1016/j.physletb.2021.136838}{{\em Phys. Lett. B}
  {\bfseries 824} (2022) 136838},
  \href{http://arxiv.org/abs/2104.10058}{{\ttfamily arXiv:2104.10058
  [hep-ph]}}.

\bibitem{Isidori:2021vtc}
G.~Isidori, D.~Lancierini, P.~Owen, and N.~Serra, ``{On the significance of new
  physics in $b\to s \ell^-\ell^+$ decays},''
  \href{http://dx.doi.org/10.1016/j.physletb.2021.136644}{{\em Phys. Lett. B}
  {\bfseries 822} (2021) 136644},
  \href{http://arxiv.org/abs/2104.05631}{{\ttfamily arXiv:2104.05631
  [hep-ph]}}.

\bibitem{Kowalska:2019ley}
K.~Kowalska, D.~Kumar, and E.~M. Sessolo, ``{Implications for new physics in
  $b\rightarrow s \mu \mu $ transitions after recent measurements by Belle and
  LHCb},'' \href{http://dx.doi.org/10.1140/epjc/s10052-019-7330-2}{{\em Eur.
  Phys. J. C} {\bfseries 79} no.~10, (2019) 840},
  \href{http://arxiv.org/abs/1903.10932}{{\ttfamily arXiv:1903.10932
  [hep-ph]}}.

\bibitem{Ciuchini:2021smi}
M.~Ciuchini, M.~Fedele, E.~Franco, A.~Paul, L.~Silvestrini, and M.~Valli,
  ``{New Physics without bias: Charming Penguins and Lepton Universality
  Violation in $b \to s \ell^+ \ell^-$ decays},''
  \href{http://arxiv.org/abs/2110.10126}{{\ttfamily arXiv:2110.10126
  [hep-ph]}}.

\bibitem{DAmico:2017mtc}
G.~D'Amico, M.~Nardecchia, P.~Panci, F.~Sannino, A.~Strumia, R.~Torre, and
  A.~Urbano, ``{Flavour anomalies after the $R_{K^*}$ measurement},''
  \href{http://dx.doi.org/10.1007/JHEP09(2017)010}{{\em JHEP} {\bfseries 09}
  (2017) 010}, \href{http://arxiv.org/abs/1704.05438}{{\ttfamily
  arXiv:1704.05438 [hep-ph]}}.

\bibitem{Super-Kamiokande:1998kpq}
{\bfseries Super-Kamiokande} Collaboration, Y.~Fukuda {\em et~al.}, ``{Evidence
  for oscillation of atmospheric neutrinos},''
  \href{http://dx.doi.org/10.1103/PhysRevLett.81.1562}{{\em Phys. Rev. Lett.}
  {\bfseries 81} (1998) 1562--1567},
  \href{http://arxiv.org/abs/hep-ex/9807003}{{\ttfamily arXiv:hep-ex/9807003}}.

\bibitem{Super-Kamiokande:2001ljr}
{\bfseries Super-Kamiokande} Collaboration, S.~Fukuda {\em et~al.}, ``{Solar
  B-8 and hep neutrino measurements from 1258 days of Super-Kamiokande data},''
  \href{http://dx.doi.org/10.1103/PhysRevLett.86.5651}{{\em Phys. Rev. Lett.}
  {\bfseries 86} (2001) 5651--5655},
  \href{http://arxiv.org/abs/hep-ex/0103032}{{\ttfamily arXiv:hep-ex/0103032}}.

\bibitem{SNO:2002tuh}
{\bfseries SNO} Collaboration, Q.~R. Ahmad {\em et~al.}, ``{Direct evidence for
  neutrino flavor transformation from neutral current interactions in the
  Sudbury Neutrino Observatory},''
  \href{http://dx.doi.org/10.1103/PhysRevLett.89.011301}{{\em Phys. Rev. Lett.}
  {\bfseries 89} (2002) 011301},
  \href{http://arxiv.org/abs/nucl-ex/0204008}{{\ttfamily
  arXiv:nucl-ex/0204008}}.

\bibitem{KamLAND:2002uet}
{\bfseries KamLAND} Collaboration, K.~Eguchi {\em et~al.}, ``{First results
  from KamLAND: Evidence for reactor anti-neutrino disappearance},''
  \href{http://dx.doi.org/10.1103/PhysRevLett.90.021802}{{\em Phys. Rev. Lett.}
  {\bfseries 90} (2003) 021802},
  \href{http://arxiv.org/abs/hep-ex/0212021}{{\ttfamily arXiv:hep-ex/0212021}}.

\bibitem{KamLAND:2004mhv}
{\bfseries KamLAND} Collaboration, T.~Araki {\em et~al.}, ``{Measurement of
  neutrino oscillation with KamLAND: Evidence of spectral distortion},''
  \href{http://dx.doi.org/10.1103/PhysRevLett.94.081801}{{\em Phys. Rev. Lett.}
  {\bfseries 94} (2005) 081801},
  \href{http://arxiv.org/abs/hep-ex/0406035}{{\ttfamily arXiv:hep-ex/0406035}}.

\bibitem{K2K:2002icj}
{\bfseries K2K} Collaboration, M.~H. Ahn {\em et~al.}, ``{Indications of
  neutrino oscillation in a 250 km long baseline experiment},''
  \href{http://dx.doi.org/10.1103/PhysRevLett.90.041801}{{\em Phys. Rev. Lett.}
  {\bfseries 90} (2003) 041801},
  \href{http://arxiv.org/abs/hep-ex/0212007}{{\ttfamily arXiv:hep-ex/0212007}}.

\bibitem{MINOS:2006foh}
{\bfseries MINOS} Collaboration, D.~G. Michael {\em et~al.}, ``{Observation of
  muon neutrino disappearance with the MINOS detectors and the NuMI neutrino
  beam},'' \href{http://dx.doi.org/10.1103/PhysRevLett.97.191801}{{\em Phys.
  Rev. Lett.} {\bfseries 97} (2006) 191801},
  \href{http://arxiv.org/abs/hep-ex/0607088}{{\ttfamily arXiv:hep-ex/0607088}}.

\bibitem{Buchmuller:1986zs}
W.~Buchmuller, R.~Ruckl, and D.~Wyler, ``{Leptoquarks in Lepton - Quark
  Collisions},'' \href{http://dx.doi.org/10.1016/S0370-2693(99)00014-3,
  10.1016/0370-2693(87)90637-X}{{\em Phys. Lett.} {\bfseries B191} (1987)
  442--448}.
[Erratum: Phys. Lett.B448,320(1999)].
%%CITATION = PHLTA,B191,442;%%.

\bibitem{Dorsner:2016wpm}
I.~Doršner, S.~Fajfer, A.~Greljo, J.~F. Kamenik, and N.~Košnik, ``{Physics of
  leptoquarks in precision experiments and at particle colliders},''
  \href{http://dx.doi.org/10.1016/j.physrep.2016.06.001}{{\em Phys. Rept.}
  {\bfseries 641} (2016) 1--68},
\href{http://arxiv.org/abs/1603.04993}{{\ttfamily arXiv:1603.04993 [hep-ph]}}.
%%CITATION = ARXIV:1603.04993;%%.

\bibitem{Peskin:1990zt}
M.~E. Peskin and T.~Takeuchi, ``{A New constraint on a strongly interacting
  Higgs sector},''
\href{http://dx.doi.org/10.1103/PhysRevLett.65.964}{{\em Phys. Rev. Lett.}
  {\bfseries 65} (1990) 964--967}.
%%CITATION = PRLTA,65,964;%%.

\bibitem{Peskin:1991sw}
M.~E. Peskin and T.~Takeuchi, ``{Estimation of oblique electroweak
  corrections},''
\href{http://dx.doi.org/10.1103/PhysRevD.46.381}{{\em Phys. Rev.} {\bfseries
  D46} (1992) 381--409}.
%%CITATION = PHRVA,D46,381;%%.

\bibitem{Grimus:2008nb}
W.~Grimus, L.~Lavoura, O.~M. Ogreid, and P.~Osland, ``{The Oblique parameters
  in multi-Higgs-doublet models},''
  \href{http://dx.doi.org/10.1016/j.nuclphysb.2008.04.019}{{\em Nucl. Phys.}
  {\bfseries B801} (2008) 81--96},
\href{http://arxiv.org/abs/0802.4353}{{\ttfamily arXiv:0802.4353 [hep-ph]}}.
%%CITATION = ARXIV:0802.4353;%%.

\bibitem{He:2001tp}
H.-J. He, N.~Polonsky, and S.-f. Su, ``{Extra families, Higgs spectrum and
  oblique corrections},''
  \href{http://dx.doi.org/10.1103/PhysRevD.64.053004}{{\em Phys. Rev. D}
  {\bfseries 64} (2001) 053004},
  \href{http://arxiv.org/abs/hep-ph/0102144}{{\ttfamily arXiv:hep-ph/0102144}}.

\bibitem{Froggatt:1991qw}
C.~D. Froggatt, R.~G. Moorhouse, and I.~G. Knowles, ``{Leading radiative
  corrections in two scalar doublet models},''
  \href{http://dx.doi.org/10.1103/PhysRevD.45.2471}{{\em Phys. Rev. D}
  {\bfseries 45} (1992) 2471--2481}.

\bibitem{Keith:1997fv}
E.~Keith and E.~Ma, ``{S, T, and leptoquarks at HERA},''
  \href{http://dx.doi.org/10.1103/PhysRevLett.79.4318}{{\em Phys. Rev. Lett.}
  {\bfseries 79} (1997) 4318--4320},
  \href{http://arxiv.org/abs/hep-ph/9707214}{{\ttfamily arXiv:hep-ph/9707214}}.

\bibitem{Dorsner:2019itg}
I.~Dor\v{s}ner, S.~Fajfer, and O.~Sumensari, ``{Muon $g-2$ and scalar
  leptoquark mixing},'' \href{http://dx.doi.org/10.1007/JHEP06(2020)089}{{\em
  JHEP} {\bfseries 06} (2020) 089},
  \href{http://arxiv.org/abs/1910.03877}{{\ttfamily arXiv:1910.03877
  [hep-ph]}}.

\bibitem{Dorsner:2020aaz}
I.~Dor\v{s}ner, S.~Fajfer, and S.~Saad, ``{$\mu \to e \gamma$ selecting scalar
  leptoquark solutions for the $(g-2)_{e,\mu}$ puzzles},''
  \href{http://dx.doi.org/10.1103/PhysRevD.102.075007}{{\em Phys. Rev. D}
  {\bfseries 102} no.~7, (2020) 075007},
  \href{http://arxiv.org/abs/2006.11624}{{\ttfamily arXiv:2006.11624
  [hep-ph]}}.

\bibitem{Crivellin:2020ukd}
A.~Crivellin, D.~M\"uller, and F.~Saturnino, ``{Leptoquarks in oblique
  corrections and Higgs signal strength: status and prospects},''
  \href{http://dx.doi.org/10.1007/JHEP11(2020)094}{{\em JHEP} {\bfseries 11}
  (2020) 094}, \href{http://arxiv.org/abs/2006.10758}{{\ttfamily
  arXiv:2006.10758 [hep-ph]}}.

\bibitem{Leveille:1977rc}
J.~P. Leveille, ``{The Second Order Weak Correction to (G-2) of the Muon in
  Arbitrary Gauge Models},''
\href{http://dx.doi.org/10.1016/0550-3213(78)90051-2}{{\em Nucl. Phys.}
  {\bfseries B137} (1978) 63--76}.
%%CITATION = NUPHA,B137,63;%%.

\bibitem{Crivellin:2018qmi}
A.~Crivellin, M.~Hoferichter, and P.~Schmidt-Wellenburg, ``{Combined
  explanations of $(g-2)_{\mu,e}$ and implications for a large muon EDM},''
  \href{http://dx.doi.org/10.1103/PhysRevD.98.113002}{{\em Phys. Rev.}
  {\bfseries D98} no.~11, (2018) 113002},
\href{http://arxiv.org/abs/1807.11484}{{\ttfamily arXiv:1807.11484 [hep-ph]}}.
%%CITATION = ARXIV:1807.11484;%%.

\bibitem{Czarnecki:2001pv}
A.~Czarnecki and W.~J. Marciano, ``{The Muon anomalous magnetic moment: A
  Harbinger for 'new physics'},''
  \href{http://dx.doi.org/10.1103/PhysRevD.64.013014}{{\em Phys. Rev. D}
  {\bfseries 64} (2001) 013014},
  \href{http://arxiv.org/abs/hep-ph/0102122}{{\ttfamily arXiv:hep-ph/0102122}}.

\bibitem{Fuyuto:2018scm}
K.~Fuyuto, M.~Ramsey-Musolf, and T.~Shen, ``{Electric Dipole Moments from
  CP-Violating Scalar Leptoquark Interactions},''
  \href{http://dx.doi.org/10.1016/j.physletb.2018.11.016}{{\em Phys. Lett. B}
  {\bfseries 788} (2019) 52--57},
  \href{http://arxiv.org/abs/1804.01137}{{\ttfamily arXiv:1804.01137
  [hep-ph]}}.

\bibitem{Bigaran:2021kmn}
I.~Bigaran and R.~R. Volkas, ``{Reflecting on Chirality: CP-violating
  extensions of the single scalar-leptoquark solutions for the $(g-2)_{e,\mu}$
  puzzles and their implications for lepton EDMs},''
  \href{http://arxiv.org/abs/2110.03707}{{\ttfamily arXiv:2110.03707
  [hep-ph]}}.

\bibitem{Borsanyi:2020mff}
S.~Borsanyi {\em et~al.}, ``{Leading hadronic contribution to the muon magnetic
  moment from lattice QCD},''
  \href{http://dx.doi.org/10.1038/s41586-021-03418-1}{{\em Nature} {\bfseries
  593} no.~7857, (2021) 51--55},
  \href{http://arxiv.org/abs/2002.12347}{{\ttfamily arXiv:2002.12347
  [hep-lat]}}.

\bibitem{Ce:2022kxy}
M.~C\`e {\em et~al.}, ``{Window observable for the hadronic vacuum polarization
  contribution to the muon $g-2$ from lattice QCD},''
  \href{http://arxiv.org/abs/2206.06582}{{\ttfamily arXiv:2206.06582
  [hep-lat]}}.

\bibitem{Alexandrou:2022amy}
C.~Alexandrou {\em et~al.}, ``{Lattice calculation of the short and
  intermediate time-distance hadronic vacuum polarization contributions to the
  muon magnetic moment using twisted-mass fermions},''
  \href{http://arxiv.org/abs/2206.15084}{{\ttfamily arXiv:2206.15084
  [hep-lat]}}.

\bibitem{RBC:2018dos}
{\bfseries RBC, UKQCD} Collaboration, T.~Blum, P.~A. Boyle, V.~G\"ulpers,
  T.~Izubuchi, L.~Jin, C.~Jung, A.~J\"uttner, C.~Lehner, A.~Portelli, and J.~T.
  Tsang, ``{Calculation of the hadronic vacuum polarization contribution to the
  muon anomalous magnetic moment},''
  \href{http://dx.doi.org/10.1103/PhysRevLett.121.022003}{{\em Phys. Rev.
  Lett.} {\bfseries 121} no.~2, (2018) 022003},
  \href{http://arxiv.org/abs/1801.07224}{{\ttfamily arXiv:1801.07224
  [hep-lat]}}.

\bibitem{Giusti:2021dvd}
D.~Giusti and S.~Simula, ``{Window contributions to the muon hadronic vacuum
  polarization with twisted-mass fermions},''
  \href{http://dx.doi.org/10.22323/1.396.0189}{{\em PoS} {\bfseries
  LATTICE2021} (2022) 189}, \href{http://arxiv.org/abs/2111.15329}{{\ttfamily
  arXiv:2111.15329 [hep-lat]}}.

\bibitem{Keshavarzi:2020bfy}
A.~Keshavarzi, W.~J. Marciano, M.~Passera, and A.~Sirlin, ``{Muon $g-2$ and
  $\Delta \alpha$ connection},''
  \href{http://dx.doi.org/10.1103/PhysRevD.102.033002}{{\em Phys. Rev. D}
  {\bfseries 102} no.~3, (2020) 033002},
  \href{http://arxiv.org/abs/2006.12666}{{\ttfamily arXiv:2006.12666
  [hep-ph]}}.

\bibitem{Crivellin:2020zul}
A.~Crivellin, M.~Hoferichter, C.~A. Manzari, and M.~Montull, ``{Hadronic Vacuum
  Polarization: $(g-2)_\mu$ versus Global Electroweak Fits},''
  \href{http://dx.doi.org/10.1103/PhysRevLett.125.091801}{{\em Phys. Rev.
  Lett.} {\bfseries 125} no.~9, (2020) 091801},
  \href{http://arxiv.org/abs/2003.04886}{{\ttfamily arXiv:2003.04886
  [hep-ph]}}.

\bibitem{DiLuzio:2021uty}
L.~Di~Luzio, A.~Masiero, P.~Paradisi, and M.~Passera, ``{New physics behind the
  new muon g-2 puzzle?},''
  \href{http://dx.doi.org/10.1016/j.physletb.2022.137037}{{\em Phys. Lett. B}
  {\bfseries 829} (2022) 137037},
  \href{http://arxiv.org/abs/2112.08312}{{\ttfamily arXiv:2112.08312
  [hep-ph]}}.

\bibitem{Ce:2022eix}
M.~C\`e, A.~G\'erardin, G.~von Hippel, H.~B. Meyer, K.~Miura, K.~Ottnad,
  A.~Risch, T.~S. Jos\'e, J.~Wilhelm, and H.~Wittig, ``{The hadronic running of
  the electromagnetic coupling and the electroweak mixing angle from lattice
  QCD},'' \href{http://arxiv.org/abs/2203.08676}{{\ttfamily arXiv:2203.08676
  [hep-lat]}}.

\bibitem{Carvunis:2021jga}
A.~Carvunis, F.~Dettori, S.~Gangal, D.~Guadagnoli, and C.~Normand, ``{On the
  effective lifetime of B$_{s}$\textrightarrow{}
  \ensuremath{\mu}\ensuremath{\mu}\ensuremath{\gamma}},''
  \href{http://dx.doi.org/10.1007/JHEP12(2021)078}{{\em JHEP} {\bfseries 12}
  (2021) 078}, \href{http://arxiv.org/abs/2102.13390}{{\ttfamily
  arXiv:2102.13390 [hep-ph]}}.

\bibitem{Dorsner:2017wwn}
I.~Dor\v{s}ner, S.~Fajfer, and N.~Ko\v{s}nik, ``{Leptoquark mechanism of
  neutrino masses within the grand unification framework},''
  \href{http://dx.doi.org/10.1140/epjc/s10052-017-4987-2}{{\em Eur. Phys. J. C}
  {\bfseries 77} no.~6, (2017) 417},
  \href{http://arxiv.org/abs/1701.08322}{{\ttfamily arXiv:1701.08322
  [hep-ph]}}.

\bibitem{Julio:2022ton}
J.~Julio, S.~Saad, and A.~Thapa, ``{A Tale of Flavor Anomalies and the Origin
  of Neutrino Mass},'' \href{http://arxiv.org/abs/2202.10479}{{\ttfamily
  arXiv:2202.10479 [hep-ph]}}.

\bibitem{Crivellin:2020tsz}
A.~Crivellin, D.~Mueller, and F.~Saturnino, ``{Correlating
  h\textrightarrow{}\ensuremath{\mu}+\ensuremath{\mu}- to the Anomalous
  Magnetic Moment of the Muon via Leptoquarks},''
  \href{http://dx.doi.org/10.1103/PhysRevLett.127.021801}{{\em Phys. Rev.
  Lett.} {\bfseries 127} no.~2, (2021) 021801},
  \href{http://arxiv.org/abs/2008.02643}{{\ttfamily arXiv:2008.02643
  [hep-ph]}}.

\bibitem{Crivellin:2020mjs}
A.~Crivellin, C.~Greub, D.~M\"uller, and F.~Saturnino, ``{Scalar Leptoquarks in
  Leptonic Processes},'' \href{http://dx.doi.org/10.1007/JHEP02(2021)182}{{\em
  JHEP} {\bfseries 02} (2021) 182},
  \href{http://arxiv.org/abs/2010.06593}{{\ttfamily arXiv:2010.06593
  [hep-ph]}}.

\bibitem{Diaz:2017lit}
B.~Diaz, M.~Schmaltz, and Y.-M. Zhong, ``{The leptoquark Hunter's guide: Pair
  production},'' \href{http://dx.doi.org/10.1007/JHEP10(2017)097}{{\em JHEP}
  {\bfseries 10} (2017) 097}, \href{http://arxiv.org/abs/1706.05033}{{\ttfamily
  arXiv:1706.05033 [hep-ph]}}.

\bibitem{Dorsner:2018ynv}
I.~Dor\v{s}ner and A.~Greljo, ``{Leptoquark toolbox for precision collider
  studies},'' \href{http://dx.doi.org/10.1007/JHEP05(2018)126}{{\em JHEP}
  {\bfseries 05} (2018) 126}, \href{http://arxiv.org/abs/1801.07641}{{\ttfamily
  arXiv:1801.07641 [hep-ph]}}.

\bibitem{ATLAS:2020dsk}
{\bfseries ATLAS} Collaboration, G.~Aad {\em et~al.}, ``{Search for pairs of
  scalar leptoquarks decaying into quarks and electrons or muons in $ \sqrt{s}
  $ = 13 TeV $pp$ collisions with the ATLAS detector},''
  \href{http://dx.doi.org/10.1007/JHEP10(2020)112}{{\em JHEP} {\bfseries 10}
  (2020) 112}, \href{http://arxiv.org/abs/2006.05872}{{\ttfamily
  arXiv:2006.05872 [hep-ex]}}.

\bibitem{ATLAS:2020xov}
{\bfseries ATLAS} Collaboration, G.~Aad {\em et~al.}, ``{Search for pair
  production of scalar leptoquarks decaying into first- or second-generation
  leptons and top quarks in proton\textendash{}proton collisions at $\sqrt{s}$
  = 13 TeV with the ATLAS detector},''
  \href{http://dx.doi.org/10.1140/epjc/s10052-021-09009-8}{{\em Eur. Phys. J.
  C} {\bfseries 81} no.~4, (2021) 313},
  \href{http://arxiv.org/abs/2010.02098}{{\ttfamily arXiv:2010.02098
  [hep-ex]}}.

\bibitem{ATLAS:2019qpq}
{\bfseries ATLAS} Collaboration, M.~Aaboud {\em et~al.}, ``{Searches for
  third-generation scalar leptoquarks in $\sqrt{s}$ = 13 TeV pp collisions with
  the ATLAS detector},'' \href{http://dx.doi.org/10.1007/JHEP06(2019)144}{{\em
  JHEP} {\bfseries 06} (2019) 144},
  \href{http://arxiv.org/abs/1902.08103}{{\ttfamily arXiv:1902.08103
  [hep-ex]}}.

\bibitem{ATLAS:2021oiz}
{\bfseries ATLAS} Collaboration, G.~Aad {\em et~al.}, ``{Search for pair
  production of third-generation scalar leptoquarks decaying into a top quark
  and a $\tau$-lepton in $pp$ collisions at $ \sqrt{s} $ = 13 TeV with the
  ATLAS detector},'' \href{http://dx.doi.org/10.1007/JHEP06(2021)179}{{\em
  JHEP} {\bfseries 06} (2021) 179},
  \href{http://arxiv.org/abs/2101.11582}{{\ttfamily arXiv:2101.11582
  [hep-ex]}}.

\bibitem{Dobrescu:2016pda}
B.~A. Dobrescu and F.~Yu, ``{Exotic Signals of Vectorlike Quarks},''
  \href{http://dx.doi.org/10.1088/1361-6471/aacbfd}{{\em J. Phys.} {\bfseries
  G45} no.~8, (2018) 08LT01},
\href{http://arxiv.org/abs/1612.01909}{{\ttfamily arXiv:1612.01909 [hep-ph]}}.
%%CITATION = ARXIV:1612.01909;%%.

\bibitem{CMS:2018zkf}
{\bfseries CMS} Collaboration, A.~M. Sirunyan {\em et~al.}, ``{Search for
  vector-like T and B quark pairs in final states with leptons at $\sqrt{s} =$
  13 TeV},'' \href{http://dx.doi.org/10.1007/JHEP08(2018)177}{{\em JHEP}
  {\bfseries 08} (2018) 177}, \href{http://arxiv.org/abs/1805.04758}{{\ttfamily
  arXiv:1805.04758 [hep-ex]}}.

\bibitem{ATLAS:2018mpo}
{\bfseries ATLAS} Collaboration, M.~Aaboud {\em et~al.}, ``{Search for pair
  production of heavy vector-like quarks decaying into high-$p_T$ $W$ bosons
  and top quarks in the lepton-plus-jets final state in $pp$ collisions at
  $\sqrt{s}=13$ TeV with the ATLAS detector},''
  \href{http://dx.doi.org/10.1007/JHEP08(2018)048}{{\em JHEP} {\bfseries 08}
  (2018) 048}, \href{http://arxiv.org/abs/1806.01762}{{\ttfamily
  arXiv:1806.01762 [hep-ex]}}.

\bibitem{Farrar:1978xj}
G.~R. Farrar and P.~Fayet, ``{Phenomenology of the Production, Decay, and
  Detection of New Hadronic States Associated with Supersymmetry},''
  \href{http://dx.doi.org/10.1016/0370-2693(78)90858-4}{{\em Phys. Lett. B}
  {\bfseries 76} (1978) 575--579}.

\bibitem{Criado:2019mvu}
J.~C. Criado and M.~Perez-Victoria, ``{Vector-like quarks with
  non-renormalizable interactions},''
  \href{http://dx.doi.org/10.1007/JHEP01(2020)057}{{\em JHEP} {\bfseries 01}
  (2020) 057}, \href{http://arxiv.org/abs/1908.08964}{{\ttfamily
  arXiv:1908.08964 [hep-ph]}}.

\bibitem{InternationalMuonCollider:2022qki}
{\bfseries International Muon Collider} Collaboration, D.~Stratakis {\em
  et~al.}, ``{A Muon Collider Facility for Physics Discovery},''
  \href{http://arxiv.org/abs/2203.08033}{{\ttfamily arXiv:2203.08033
  [physics.acc-ph]}}.

\bibitem{Jindariani:2022gxj}
S.~Jindariani {\em et~al.}, ``{Promising Technologies and R\&D Directions for
  the Future Muon Collider Detectors},'' in {\em {2022 Snowmass Summer Study}}.
\newblock 3, 2022.
\newblock \href{http://arxiv.org/abs/2203.07224}{{\ttfamily arXiv:2203.07224
  [physics.ins-det]}}.

\bibitem{Aime:2022flm}
C.~Aime {\em et~al.}, ``{Muon Collider Physics Summary},''
  \href{http://arxiv.org/abs/2203.07256}{{\ttfamily arXiv:2203.07256
  [hep-ph]}}.

\bibitem{MuonCollider:2022xlm}
{\bfseries Muon Collider} Collaboration, J.~de~Blas {\em et~al.}, ``{The
  physics case of a 3 TeV muon collider stage},''
  \href{http://arxiv.org/abs/2203.07261}{{\ttfamily arXiv:2203.07261
  [hep-ph]}}.

\bibitem{FCC:2018byv}
{\bfseries FCC} Collaboration, A.~Abada {\em et~al.}, ``{FCC Physics
  Opportunities}: {Future Circular Collider Conceptual Design Report Volume
  1},'' \href{http://dx.doi.org/10.1140/epjc/s10052-019-6904-3}{{\em Eur. Phys.
  J. C} {\bfseries 79} no.~6, (2019) 474}.

\bibitem{FCC:2018vvp}
{\bfseries FCC} Collaboration, A.~Abada {\em et~al.}, ``{FCC-hh: The Hadron
  Collider}: {Future Circular Collider Conceptual Design Report Volume 3},''
  \href{http://dx.doi.org/10.1140/epjst/e2019-900087-0}{{\em Eur. Phys. J. ST}
  {\bfseries 228} no.~4, (2019) 755--1107}.

\bibitem{Bernardi:2022hny}
G.~Bernardi {\em et~al.}, ``{The Future Circular Collider: a Summary for the US
  2021 Snowmass Process},'' \href{http://arxiv.org/abs/2203.06520}{{\ttfamily
  arXiv:2203.06520 [hep-ex]}}.

\bibitem{DiLuzio:2017chi}
L.~Di~Luzio and M.~Nardecchia, ``{What is the scale of new physics behind the
  $B$-flavour anomalies?},''
  \href{http://dx.doi.org/10.1140/epjc/s10052-017-5118-9}{{\em Eur. Phys. J.}
  {\bfseries C77} no.~8, (2017) 536},
\href{http://arxiv.org/abs/1706.01868}{{\ttfamily arXiv:1706.01868 [hep-ph]}}.
%%CITATION = ARXIV:1706.01868;%%.

\bibitem{Azatov:2022itm}
A.~Azatov, F.~Garosi, A.~Greljo, D.~Marzocca, J.~Salko, and S.~Trifinopoulos,
  ``{New Physics in $b \to s \mu \mu$: FCC-hh or a Muon Collider?},''
  \href{http://arxiv.org/abs/2205.13552}{{\ttfamily arXiv:2205.13552
  [hep-ph]}}.

\bibitem{Baker:2021yli}
M.~J. Baker, P.~Cox, and R.~R. Volkas, ``{Radiative muon mass models and
  $(g-2)_\mu$},'' \href{http://dx.doi.org/10.1007/JHEP05(2021)174}{{\em JHEP}
  {\bfseries 05} (2021) 174}, \href{http://arxiv.org/abs/2103.13401}{{\ttfamily
  arXiv:2103.13401 [hep-ph]}}.

\bibitem{CMS:2020xwi}
{\bfseries CMS} Collaboration, A.~M. Sirunyan {\em et~al.}, ``{Evidence for
  Higgs boson decay to a pair of muons},''
  \href{http://dx.doi.org/10.1007/JHEP01(2021)148}{{\em JHEP} {\bfseries 01}
  (2021) 148}, \href{http://arxiv.org/abs/2009.04363}{{\ttfamily
  arXiv:2009.04363 [hep-ex]}}.

\bibitem{ATLAS:2020fzp}
{\bfseries ATLAS} Collaboration, G.~Aad {\em et~al.}, ``{A search for the
  dimuon decay of the Standard Model Higgs boson with the ATLAS detector},''
  \href{http://dx.doi.org/10.1016/j.physletb.2020.135980}{{\em Phys. Lett. B}
  {\bfseries 812} (2021) 135980},
  \href{http://arxiv.org/abs/2007.07830}{{\ttfamily arXiv:2007.07830
  [hep-ex]}}.

\bibitem{Saad:2020ucl}
S.~Saad and A.~Thapa, ``{Common origin of neutrino masses and $R_{D^{(\ast)}}$,
  $R_{K^{(\ast)}}$ anomalies},''
  \href{http://dx.doi.org/10.1103/PhysRevD.102.015014}{{\em Phys. Rev. D}
  {\bfseries 102} no.~1, (2020) 015014},
  \href{http://arxiv.org/abs/2004.07880}{{\ttfamily arXiv:2004.07880
  [hep-ph]}}.

\bibitem{Saad:2020ihm}
S.~Saad, ``{Combined explanations of $(g-2)_{\mu}$, $R_{D^{(*)}}$,
  $R_{K^{(*)}}$ anomalies in a two-loop radiative neutrino mass model},''
  \href{http://dx.doi.org/10.1103/PhysRevD.102.015019}{{\em Phys. Rev. D}
  {\bfseries 102} no.~1, (2020) 015019},
  \href{http://arxiv.org/abs/2005.04352}{{\ttfamily arXiv:2005.04352
  [hep-ph]}}.

\bibitem{Julio:2022bue}
J.~Julio, S.~Saad, and A.~Thapa, ``{Marriage between neutrino mass and flavor
  anomalies},'' \href{http://arxiv.org/abs/2203.15499}{{\ttfamily
  arXiv:2203.15499 [hep-ph]}}.

\bibitem{Pas:2015hca}
H.~Päs and E.~Schumacher, ``{Common origin of $R_K$ and neutrino masses},''
  \href{http://dx.doi.org/10.1103/PhysRevD.92.114025}{{\em Phys. Rev.}
  {\bfseries D92} no.~11, (2015) 114025},
\href{http://arxiv.org/abs/1510.08757}{{\ttfamily arXiv:1510.08757 [hep-ph]}}.
%%CITATION = ARXIV:1510.08757;%%.

\bibitem{Cheung:2016fjo}
K.~Cheung, T.~Nomura, and H.~Okada, ``{Testable radiative neutrino mass model
  without additional symmetries and explanation for the $b \to s \ell^+ \ell^-$
  anomaly},'' \href{http://dx.doi.org/10.1103/PhysRevD.94.115024}{{\em Phys.
  Rev. D} {\bfseries 94} no.~11, (2016) 115024},
  \href{http://arxiv.org/abs/1610.02322}{{\ttfamily arXiv:1610.02322
  [hep-ph]}}.

\bibitem{Deppisch:2016qqd}
F.~F. Deppisch, S.~Kulkarni, H.~Päs, and E.~Schumacher, ``{Leptoquark patterns
  unifying neutrino masses, flavor anomalies, and the diphoton excess},''
  \href{http://dx.doi.org/10.1103/PhysRevD.94.013003}{{\em Phys. Rev.}
  {\bfseries D94} no.~1, (2016) 013003},
\href{http://arxiv.org/abs/1603.07672}{{\ttfamily arXiv:1603.07672 [hep-ph]}}.
%%CITATION = ARXIV:1603.07672;%%.

\bibitem{Cai:2017wry}
Y.~Cai, J.~Gargalionis, M.~A. Schmidt, and R.~R. Volkas, ``{Reconsidering the
  One Leptoquark solution: flavor anomalies and neutrino mass},''
  \href{http://dx.doi.org/10.1007/JHEP10(2017)047}{{\em JHEP} {\bfseries 10}
  (2017) 047},
\href{http://arxiv.org/abs/1704.05849}{{\ttfamily arXiv:1704.05849 [hep-ph]}}.
%%CITATION = ARXIV:1704.05849;%%.

\bibitem{Guo:2017gxp}
S.-Y. Guo, Z.-L. Han, B.~Li, Y.~Liao, and X.-D. Ma, ``{Interpreting the
  $R_{K^{(*)}}$ anomaly in the colored Zee–Babu model},''
  \href{http://dx.doi.org/10.1016/j.nuclphysb.2018.01.024}{{\em Nucl. Phys.}
  {\bfseries B928} (2018) 435--447},
\href{http://arxiv.org/abs/1707.00522}{{\ttfamily arXiv:1707.00522 [hep-ph]}}.
%%CITATION = ARXIV:1707.00522;%%.

\bibitem{Hati:2018fzc}
C.~Hati, G.~Kumar, J.~Orloff, and A.~M. Teixeira, ``{Reconciling $B$-meson
  decay anomalies with neutrino masses, dark matter and constraints from
  flavour violation},'' \href{http://dx.doi.org/10.1007/JHEP11(2018)011}{{\em
  JHEP} {\bfseries 11} (2018) 011},
\href{http://arxiv.org/abs/1806.10146}{{\ttfamily arXiv:1806.10146 [hep-ph]}}.
%%CITATION = ARXIV:1806.10146;%%.

\bibitem{Datta:2019tuj}
A.~Datta, D.~Sachdeva, and J.~Waite, ``{Unified explanation of $b \to s \mu^+
  \mu^-$ anomalies, neutrino masses, and $B\rightarrow \pi K$ puzzle},''
  \href{http://dx.doi.org/10.1103/PhysRevD.100.055015}{{\em Phys. Rev.}
  {\bfseries D100} no.~5, (2019) 055015},
\href{http://arxiv.org/abs/1905.04046}{{\ttfamily arXiv:1905.04046 [hep-ph]}}.
%%CITATION = ARXIV:1905.04046;%%.

\bibitem{Popov:2019tyc}
O.~Popov, M.~A. Schmidt, and G.~White, ``{$R_2$ as a single leptoquark solution
  to $R_{D^{(*)}}$ and $R_{K^{(*)}}$},''
  \href{http://dx.doi.org/10.1103/PhysRevD.100.035028}{{\em Phys. Rev.}
  {\bfseries D100} no.~3, (2019) 035028},
\href{http://arxiv.org/abs/1905.06339}{{\ttfamily arXiv:1905.06339 [hep-ph]}}.
%%CITATION = ARXIV:1905.06339;%%.

\bibitem{Dev:2020qet}
P.~B. Dev, R.~Mohanta, S.~Patra, and S.~Sahoo, ``{Unified explanation of flavor
  anomalies, radiative neutrino mass and ANITA anomalous events in a vector
  leptoquark model},'' \href{http://arxiv.org/abs/2004.09464}{{\ttfamily
  arXiv:2004.09464 [hep-ph]}}.

\bibitem{Babu:2020hun}
K.~S. Babu, P.~S.~B. Dev, S.~Jana, and A.~Thapa, ``{Unified framework for
  $B$-anomalies, muon $g − 2$ and neutrino masses},''
  \href{http://dx.doi.org/10.1007/JHEP03(2021)179}{{\em JHEP} {\bfseries 03}
  (2021) 179}, \href{http://arxiv.org/abs/2009.01771}{{\ttfamily
  arXiv:2009.01771 [hep-ph]}}.

\bibitem{BaBar:2009hkt}
{\bfseries BaBar} Collaboration, B.~Aubert {\em et~al.}, ``{Searches for Lepton
  Flavor Violation in the Decays tau+- ---\ensuremath{>} e+- gamma and tau+-
  ---\ensuremath{>} mu+- gamma},''
  \href{http://dx.doi.org/10.1103/PhysRevLett.104.021802}{{\em Phys. Rev.
  Lett.} {\bfseries 104} (2010) 021802},
  \href{http://arxiv.org/abs/0908.2381}{{\ttfamily arXiv:0908.2381 [hep-ex]}}.

\bibitem{Aushev:2010bq}
T.~Aushev {\em et~al.}, ``{Physics at Super B Factory},''
  \href{http://arxiv.org/abs/1002.5012}{{\ttfamily arXiv:1002.5012 [hep-ex]}}.

\bibitem{Belle:2017oht}
{\bfseries Belle} Collaboration, J.~Grygier {\em et~al.}, ``{Search for
  $\boldsymbol{B\to h\nu\bar{\nu}}$ decays with semileptonic tagging at
  Belle},'' \href{http://dx.doi.org/10.1103/PhysRevD.96.091101}{{\em Phys. Rev.
  D} {\bfseries 96} no.~9, (2017) 091101},
  \href{http://arxiv.org/abs/1702.03224}{{\ttfamily arXiv:1702.03224
  [hep-ex]}}. [Addendum: Phys.Rev.D 97, 099902 (2018)].

\bibitem{Belle-II:2018jsg}
{\bfseries Belle-II} Collaboration, W.~Altmannshofer {\em et~al.}, ``{The Belle
  II Physics Book},'' \href{http://dx.doi.org/10.1093/ptep/ptz106}{{\em PTEP}
  {\bfseries 2019} no.~12, (2019) 123C01},
  \href{http://arxiv.org/abs/1808.10567}{{\ttfamily arXiv:1808.10567
  [hep-ex]}}. [Erratum: PTEP 2020, 029201 (2020)].

\bibitem{Georgi:1974sy}
H.~Georgi and S.~L. Glashow, ``{Unity of All Elementary Particle Forces},''
\href{http://dx.doi.org/10.1103/PhysRevLett.32.438}{{\em Phys. Rev. Lett.}
  {\bfseries 32} (1974) 438--441}.
%%CITATION = PRLTA,32,438;%%.

\bibitem{Georgi:1979df}
H.~Georgi and C.~Jarlskog, ``{A New Lepton - Quark Mass Relation in a Unified
  Theory},'' \href{http://dx.doi.org/10.1016/0370-2693(79)90842-6}{{\em Phys.
  Lett. B} {\bfseries 86} (1979) 297--300}.

\bibitem{Saad:2019vjo}
S.~Saad, ``{Origin of a two-loop neutrino mass from SU(5) grand unification},''
  \href{http://dx.doi.org/10.1103/PhysRevD.99.115016}{{\em Phys. Rev.}
  {\bfseries D99} no.~11, (2019) 115016},
\href{http://arxiv.org/abs/1902.11254}{{\ttfamily arXiv:1902.11254 [hep-ph]}}.
%%CITATION = ARXIV:1902.11254;%%.

\bibitem{Lavoura:2003xp}
L.~Lavoura, ``{General formulae for $f(1) \to f(2) \gamma$},''
  \href{http://dx.doi.org/10.1140/epjc/s2003-01212-7}{{\em Eur. Phys. J.}
  {\bfseries C29} (2003) 191--195},
\href{http://arxiv.org/abs/hep-ph/0302221}{{\ttfamily arXiv:hep-ph/0302221
  [hep-ph]}}.
%%CITATION = HEP-PH/0302221;%%.

\bibitem{TheMEG:2016wtm}
{\bfseries MEG} Collaboration, A.~M. Baldini {\em et~al.}, ``{Search for the
  lepton flavour violating decay $\mu ^+ \rightarrow \mathrm {e}^+ \gamma $
  with the full dataset of the MEG experiment},''
  \href{http://dx.doi.org/10.1140/epjc/s10052-016-4271-x}{{\em Eur. Phys. J.}
  {\bfseries C76} no.~8, (2016) 434},
\href{http://arxiv.org/abs/1605.05081}{{\ttfamily arXiv:1605.05081 [hep-ex]}}.
%%CITATION = ARXIV:1605.05081;%%.

\bibitem{Aubert:2009ag}
{\bfseries BaBar} Collaboration, B.~Aubert {\em et~al.}, ``{Searches for Lepton
  Flavor Violation in the Decays $\tau^{\pm} \to e^{\pm} \gamma$ and
  $\tau^{\pm} \to \mu^{\pm} \gamma$},''
  \href{http://dx.doi.org/10.1103/PhysRevLett.104.021802}{{\em Phys. Rev.
  Lett.} {\bfseries 104} (2010) 021802},
\href{http://arxiv.org/abs/0908.2381}{{\ttfamily arXiv:0908.2381 [hep-ex]}}.
%%CITATION = ARXIV:0908.2381;%%.

\bibitem{Tanabashi:2018oca}
{\bfseries Particle Data Group} Collaboration, M.~Tanabashi {\em et~al.},
  ``{Review of Particle Physics},''
\href{http://dx.doi.org/10.1103/PhysRevD.98.030001}{{\em Phys. Rev.} {\bfseries
  D98} no.~3, (2018) 030001}.
%%CITATION = PHRVA,D98,030001;%%.

\bibitem{Arnan:2019olv}
P.~Arnan, D.~Becirevic, F.~Mescia, and O.~Sumensari, ``{Probing low energy
  scalar leptoquarks by the leptonic $W$ and $Z$ couplings},''
  \href{http://dx.doi.org/10.1007/JHEP02(2019)109}{{\em JHEP} {\bfseries 02}
  (2019) 109},
\href{http://arxiv.org/abs/1901.06315}{{\ttfamily arXiv:1901.06315 [hep-ph]}}.
%%CITATION = ARXIV:1901.06315;%%.

\bibitem{ALEPH:2005ab}
{\bfseries ALEPH, DELPHI, L3, OPAL, SLD, LEP Electroweak Working Group, SLD
  Electroweak Group, SLD Heavy Flavour Group} Collaboration, S.~Schael {\em
  et~al.}, ``{Precision electroweak measurements on the $Z$ resonance},''
  \href{http://dx.doi.org/10.1016/j.physrep.2005.12.006}{{\em Phys. Rept.}
  {\bfseries 427} (2006) 257--454},
  \href{http://arxiv.org/abs/hep-ex/0509008}{{\ttfamily arXiv:hep-ex/0509008}}.

\bibitem{Kitano:2002mt}
R.~Kitano, M.~Koike, and Y.~Okada, ``{Detailed calculation of lepton flavor
  violating muon electron conversion rate for various nuclei},''
  \href{http://dx.doi.org/10.1103/PhysRevD.76.059902,
  10.1103/PhysRevD.66.096002}{{\em Phys. Rev.} {\bfseries D66} (2002) 096002},
  \href{http://arxiv.org/abs/hep-ph/0203110}{{\ttfamily arXiv:hep-ph/0203110
  [hep-ph]}}.
[Erratum: Phys. Rev.D76,059902(2007)].
%%CITATION = HEP-PH/0203110;%%.

\bibitem{Bertl:2006up}
{\bfseries SINDRUM II} Collaboration, W.~H. Bertl {\em et~al.}, ``{A Search for
  muon to electron conversion in muonic gold},''
\href{http://dx.doi.org/10.1140/epjc/s2006-02582-x}{{\em Eur. Phys. J.}
  {\bfseries C47} (2006) 337--346}.
%%CITATION = EPHJA,C47,337;%%.

\bibitem{Kurup:2011zza}
{\bfseries COMET} Collaboration, A.~Kurup, ``{The COherent Muon to Electron
  Transition (COMET) experiment},''
\href{http://dx.doi.org/10.1016/j.nuclphysbps.2011.06.008}{{\em Nucl. Phys.
  Proc. Suppl.} {\bfseries 218} (2011) 38--43}.
%%CITATION = NUPHZ,218,38;%%.

\bibitem{Cui:2009zz}
{\bfseries COMET} Collaboration, Y.~G. Cui {\em et~al.},
``{Conceptual design report for experimental search for lepton flavor violating
  mu- e- conversion at sensitivity of 10**(-16) with a slow-extracted bunched
  proton beam (COMET)},''.
%%CITATION = KEK-2009-10;%%.

\bibitem{Chang:2000ac}
C.-H. Chang, S.-L. Chen, T.-F. Feng, and X.-Q. Li, ``{The Lifetime of $B_c$
  meson and some relevant problems},''
  \href{http://dx.doi.org/10.1103/PhysRevD.64.014003}{{\em Phys. Rev.}
  {\bfseries D64} (2001) 014003},
\href{http://arxiv.org/abs/hep-ph/0007162}{{\ttfamily arXiv:hep-ph/0007162
  [hep-ph]}}.
%%CITATION = HEP-PH/0007162;%%.

\bibitem{Adamov:2018vin}
{\bfseries COMET} Collaboration, R.~Abramishvili {\em et~al.}, ``{COMET Phase-I
  Technical Design Report},''
\href{http://arxiv.org/abs/1812.09018}{{\ttfamily arXiv:1812.09018
  [physics.ins-det]}}.
%%CITATION = ARXIV:1812.09018;%%.

\bibitem{Bartoszek:2014mya}
{\bfseries Mu2e} Collaboration, L.~Bartoszek {\em et~al.}, ``{Mu2e Technical
  Design Report},''
\href{http://arxiv.org/abs/1501.05241}{{\ttfamily arXiv:1501.05241
  [physics.ins-det]}}.
%%CITATION = ARXIV:1501.05241;%%.

\bibitem{Pezzullo:2018fzp}
{\bfseries Mu2e} Collaboration, G.~Pezzullo, ``{Mu2e: A Search for Charged
  Lepton Flavor Violation in $\rm \mu N \to e N$ Conversion with a Sensitivity
  $< 10^{-16}$},''
\href{http://dx.doi.org/10.22323/1.340.0583}{{\em PoS} {\bfseries ICHEP2018}
  (2019) 583}.
%%CITATION = POSCI,ICHEP2018,583;%%.

\bibitem{Bonventre:2019grv}
{\bfseries Mu2e} Collaboration, R.~Bonventre, ``{Searching for muon to electron
  conversion: The Mu2e experiment at Fermilab},''
\href{http://dx.doi.org/10.21468/SciPostPhysProc.1.038}{{\em SciPost Phys.
  Proc.} {\bfseries 1} (2019) 038}.
%%CITATION = INSPIRE-1729457;%%.

\bibitem{Buras:2014fpa}
A.~J. Buras, J.~Girrbach-Noe, C.~Niehoff, and D.~M. Straub, ``{$ B\to
  {K}^{\left(\ast \right)}\nu \overline{\nu} $ decays in the Standard Model and
  beyond},'' \href{http://dx.doi.org/10.1007/JHEP02(2015)184}{{\em JHEP}
  {\bfseries 02} (2015) 184},
\href{http://arxiv.org/abs/1409.4557}{{\ttfamily arXiv:1409.4557 [hep-ph]}}.
%%CITATION = ARXIV:1409.4557;%%.

\bibitem{Grygier:2017tzo}
{\bfseries Belle} Collaboration, J.~Grygier {\em et~al.}, ``{Search for
  $\boldsymbol{B\to h\nu\bar{\nu}}$ decays with semileptonic tagging at
  Belle},'' \href{http://dx.doi.org/10.1103/PhysRevD.96.091101}{{\em Phys. Rev.
  D} {\bfseries 96} no.~9, (2017) 091101},
  \href{http://arxiv.org/abs/1702.03224}{{\ttfamily arXiv:1702.03224
  [hep-ex]}}. [Addendum: Phys.Rev.D 97, 099902 (2018)].

\bibitem{Marzocca:2018wcf}
D.~Marzocca, ``{Addressing the B-physics anomalies in a fundamental Composite
  Higgs Model},'' \href{http://dx.doi.org/10.1007/JHEP07(2018)121}{{\em JHEP}
  {\bfseries 07} (2018) 121},
\href{http://arxiv.org/abs/1803.10972}{{\ttfamily arXiv:1803.10972 [hep-ph]}}.
%%CITATION = ARXIV:1803.10972;%%.

\bibitem{Lenz:2010gu}
A.~Lenz, U.~Nierste, J.~Charles, S.~Descotes-Genon, A.~Jantsch, C.~Kaufhold,
  H.~Lacker, S.~Monteil, V.~Niess, and S.~T'Jampens, ``{Anatomy of New Physics
  in $B - \bar{B}$ mixing},''
  \href{http://dx.doi.org/10.1103/PhysRevD.83.036004}{{\em Phys. Rev. D}
  {\bfseries 83} (2011) 036004},
  \href{http://arxiv.org/abs/1008.1593}{{\ttfamily arXiv:1008.1593 [hep-ph]}}.

\bibitem{Bobeth:2017ecx}
C.~Bobeth and A.~J. Buras, ``{Leptoquarks meet $\varepsilon'/\varepsilon$ and
  rare Kaon processes},'' \href{http://dx.doi.org/10.1007/JHEP02(2018)101}{{\em
  JHEP} {\bfseries 02} (2018) 101},
\href{http://arxiv.org/abs/1712.01295}{{\ttfamily arXiv:1712.01295 [hep-ph]}}.
%%CITATION = ARXIV:1712.01295;%%.

\bibitem{Crivellin:2019dwb}
A.~Crivellin, D.~Müller, and F.~Saturnino, ``{Flavor Phenomenology of the
  Leptoquark Singlet-Triplet Model},''
\href{http://arxiv.org/abs/1912.04224}{{\ttfamily arXiv:1912.04224 [hep-ph]}}.
%%CITATION = ARXIV:1912.04224;%%.

\bibitem{Bona:2006sa}
{\bfseries UTfit} Collaboration, M.~Bona {\em et~al.}, ``{Constraints on new
  physics from the quark mixing unitarity triangle},''
  \href{http://dx.doi.org/10.1103/PhysRevLett.97.151803}{{\em Phys. Rev. Lett.}
  {\bfseries 97} (2006) 151803},
  \href{http://arxiv.org/abs/hep-ph/0605213}{{\ttfamily arXiv:hep-ph/0605213}}.

\bibitem{Jubb:2016mvq}
T.~Jubb, M.~Kirk, A.~Lenz, and G.~Tetlalmatzi-Xolocotzi, ``{On the ultimate
  precision of meson mixing observables},''
  \href{http://dx.doi.org/10.1016/j.nuclphysb.2016.12.020}{{\em Nucl. Phys. B}
  {\bfseries 915} (2017) 431--453},
  \href{http://arxiv.org/abs/1603.07770}{{\ttfamily arXiv:1603.07770
  [hep-ph]}}.

\bibitem{Bona:2008jn}
{\bfseries UTfit} Collaboration, M.~Bona {\em et~al.}, ``{First Evidence of New
  Physics in $b \longleftrightarrow s$ Transitions},''
  \href{http://dx.doi.org/10.1186/1754-0410-3-6}{{\em PMC Phys. A} {\bfseries
  3} (2009) 6}, \href{http://arxiv.org/abs/0803.0659}{{\ttfamily
  arXiv:0803.0659 [hep-ph]}}.

\bibitem{UTfit:2007eik}
{\bfseries UTfit} Collaboration, M.~Bona {\em et~al.}, ``{Model-independent
  constraints on $\Delta F=2$ operators and the scale of new physics},''
  \href{http://dx.doi.org/10.1088/1126-6708/2008/03/049}{{\em JHEP} {\bfseries
  03} (2008) 049}, \href{http://arxiv.org/abs/0707.0636}{{\ttfamily
  arXiv:0707.0636 [hep-ph]}}.

\end{thebibliography}\endgroup
\end{document}